\documentclass{book}
\usepackage{amsmath, amsthm, amssymb}
\usepackage{epsfig}
\newcommand{\al}{\alpha}

\newcommand{\nts}{\negthickspace}
\author{Claire Levaillant}
\date{\textit{Work with the Microsoft Research Station Q team}}
\title{Topological quantum computation within the anyonic system the Kauffman-Jones version of $SU(2)$ Chern-Simons theory at level $4$}
\begin{document}
\maketitle
\let\cleardoublepage\clearpage
\chapter*{Introduction}
We present protocols allowing to perform any unitary operation on $n$-qubits on a topological quantum computer within the anyonic system the Kauffman-Jones version of $SU(2)_4$. The minor modifications resulting from studying $SU(2)_4$ instead of the Kauffman-Jones version of $SU(2)_4$ get explained in \cite{Qet}. The quantum gates are made by braiding and measuring quasiparticles called anyons. \\

The first two chapters deal with $1$-qudit gates for $d=2,3$, and the last chapter deals with $2$-qubit gates.
In $2001$, Ranee and Jean-Luc Brylinski showed as part of their work that if we can approximate any $1$-qubit gate and if we have only one $2$-qubit entangling gate, then we can approximate any $n$-qubit gate.\\

The goal of our paper is to tell how to braid and to measure the particles in order to make a few gates (a few $1$-qubit gates and at least one $2$-qubit entangling gate) from which one can approximate any other gate.\\

By braiding only, we obtain only finitely many $1$-qubit gates. But, by adding measurement based operations, we are able to make an irrational phase gate, which makes the group of gates issued from braiding become infinite and dense in $SU(2)$ (overall phases do not affect quantum computation).
The gate is made from an ancilla which we prepare so that it has an irrational relative phase. Making ancillas is simpler than making gates and is quite comfortable because any measurement operation is allowed. Indeed, in case of a bad measurement outcome, we can simply dispose of the ancilla. Also, when making quantum gates, a big challenge of measurement based operations is that the effect of the measurement on the system should be independent from the input. This is a difficult condition to have. In order to bypass this difficulty, we prepare an ancilla by braiding and measurement and then fuse the ancilla into the input in order to form the gate. The difficulty has however not completely vanished. In the process we must do measurements. And so we must have adequate ancillas so that no quantum information gets lost, nor distorted during the fusion of the ancilla  with the input. Making such ancillas is a priori a challenge but is still a more reasonable challenge than making a gate directly, due to the large number of possibilities offered in terms of braiding, measurements, number of particles used.\\

The third  and last chapter deals with realizing all the $2$-qubit permutation gates, some of which are entangling gates (like the CNOT gate).
The chapter also relies on other published protocols which are not part of this paper.

\chapter{From ancilla to quantum gate}
\begin{center}Note\end{center}
This chapter grew out of many inspiring discussions with Michael Freedman and is the first stone in a series of two chapters leading to the production of irrational phase gates in the Kauffman-Jones version of $SU(2)$ Chern-Simons theory at level $4$.
\begin{center} Abstract \end{center}
We provide a way to turn an ancilla into a gate for specific ancillas in the Kauffman-Jones version of $SU(2)$ Chern-Simons theory at level $4$. We deal with both the qubit $1221$ and the qutrit $2222$. Together with ancilla preparations, our protocols are later used to make irrational phase gates, leading to universal single qubit and qutrit gates.
\section{Motivation of the work}
The framework of this work is the Kauffman-Jones version of $SU(2)$ Chern-Simons theory at level $4$. In this theory, there are five particle types, called anyons, with respective topological charges $0$, $1$, $2$, $3$ and $4$. The anyons obey to fusion rules that are governed by
$$\left\lbrace\begin{array}{l}
a\leq b+c,\,b\leq c+a,\,c\leq a+b\\
a+b+c\leq 8\\
a+b+c\,\;\text{is even}\end{array}\right.,$$
where $a$, $b$ and $c$ denote the topological charges of the particles.
The operations we do consist of braiding the anyons, fusing them and measuring them.
Some measurements are fusion measurements and other measurements are done by interferometry, see \cite{BO}, \cite{BO2}, \cite{BO3}.
For basic facts about recoupling theory, we refer the reader to \cite{KL}, except the theory which we use is unitary. In particular we deal with unitary theta symbols and unitary $6j$-symbols (see \cite{ZW} and Appendix of \cite{CL}). The value of the Kauffman constant is, using the same notations as in \cite{KL},
$$A=i\,e^{-i\frac{\pi}{12}}$$
The main four moves which we use throughout the paper are summarized below.
\begin{itemize}
\item The "$F$-move"
\begin{center}
\epsfig{file=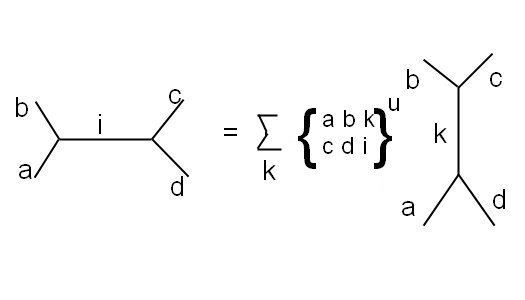, height=5cm}
\end{center}
The brackets are called unitary $6j$-symbols.
\item The "$R$-move"
\begin{center}
\epsfig{file=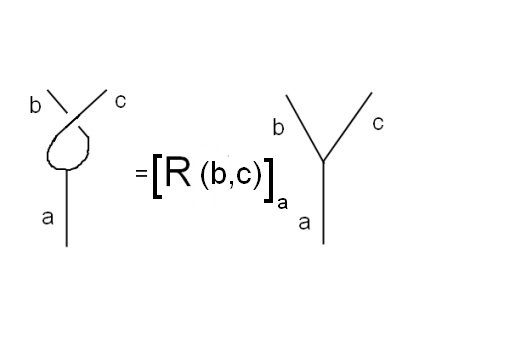, height=5cm}
\end{center}
\item The "theta move"
\begin{center}
\epsfig{file=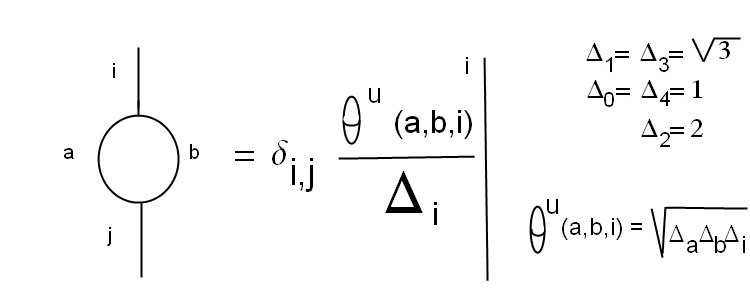, height=5cm}
\end{center}
The delta's are the quantum dimensions and the $\theta^u$ are the unitary theta symbols.
\item The "bubble removal"
\begin{center}
\epsfig{file=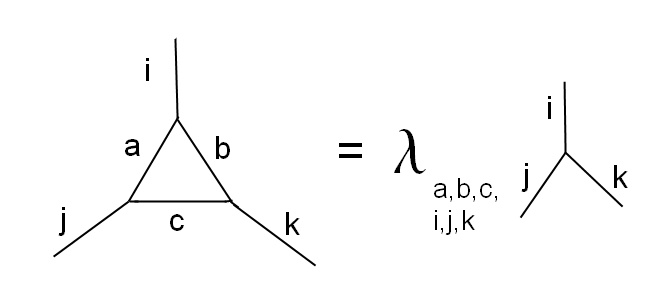, height=5cm}
\end{center}
The coefficient is obtained by doing an F-move, followed by a Theta move.
Note that $i$, $j$ and $k$ must obey to the fusion rules, else the evaluation is zero.
\end{itemize}

We study here two anyonic systems: the quantum bit formed by four anyons of topological charges $1221$ and the quantum trit formed by four anyons of topological charges $2222$. \begin{center}
\epsfig{file=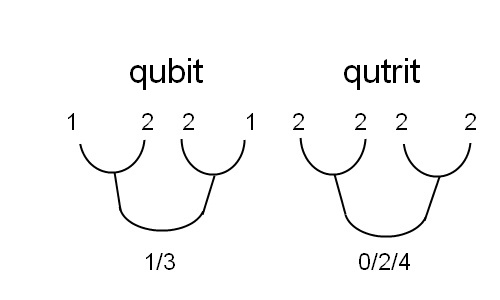, height=6cm}\end{center}

We show in each case how, starting with adequate ancillas, we are able to turn them into quantum gates. We got inspired by \cite{BK}. The main idea is to fuse the ancilla into the input and avoid leakage of quantum information during the process by imposing some conditions on the ancillas we use. Inventing protocols that produce interesting ancillas is much easier than inventing protocols that directly produce interesting quantum gates. Indeed, when doing measurements during the ancilla production, if we don't measure the precise outcome we wish to measure, then we can just dispose of the ancilla. This allows to not worry about recoveries for bad measurement outcomes. And these recoveries can be challenging. By the fact that the range of allowed possibilities is considerably enhanced since any measurement is permitted, it is possible to make ancillas with interesting amplitude ratios and interesting relative phases. \\\\ \textbf{In what follows, the ancillas and the gates are all defined up to some overall complex scalars.} \\\\The method which we expose in this paper is the following.
\begin{itemize}
\item For the qubit $1221$, given an ancilla
$$a\,|1> + b\,|3>$$ where $a$ and $b$ are two complex numbers of equal norms, how to make the gate
$$\begin{pmatrix}
a&0\\0&b
\end{pmatrix}$$
\item For the qutrit $2222$, given an ancilla $$|0>+\sqrt{2}\,e^{i\,\al}\,|2>+|4>$$
    where $\al$ is any phase, how to make the gate
    $$\begin{pmatrix}
    1&0&0\\
    0&e^{i\,\al}&0\\
    0&0&1\end{pmatrix}$$
\end{itemize}
The protocols above are later used in Chapter $2$ to make infinite order phase gates for the qubit $1221$ and the qutrit $2222$, after we took care of producing adequate ancillas. Thus leading to universal single qubit and qutrit gates by only three operations consisting of braiding, measuring and fusing particles.
\section{The qubit $1221$}
The anyons will be numbered from left to right. We bring the input and the ancilla close to each other and do an F-move like on the picture below. \vspace{-0.7cm}
\begin{center}
\epsfig{file=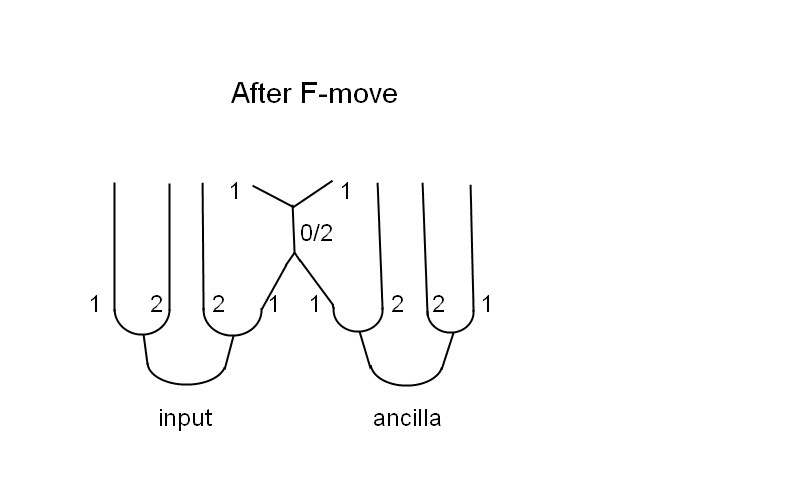, height=8cm}
\end{center}
We measure anyons $4$ and $5$. If the measurement outcome is $0$, then we have fused anyons $4$ and $5$ and we have a pair of $1$'s which we dispose of. \\
\epsfig{file=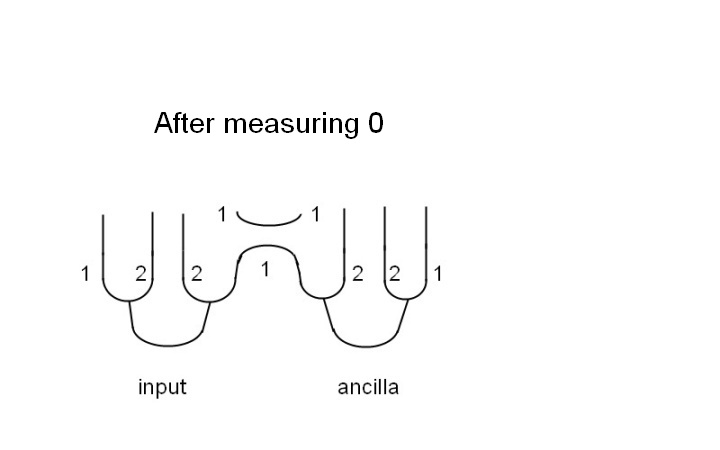, height=8cm}
If we rather measure $2$, then we do an F-move on the edge labeled $2$ of the measurement (no physical action here, only for the clarity of the presentation) and we measure the last four anyons by interferometry like on the figure below. \begin{center}
\epsfig{file=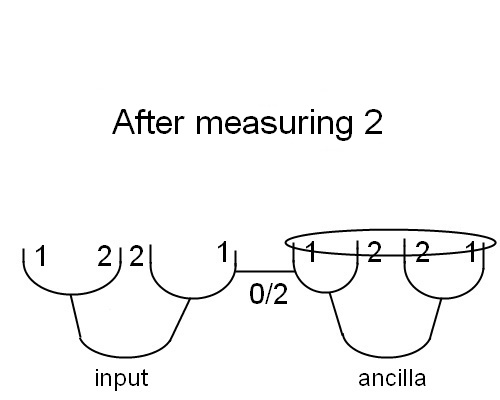, height=6cm}\end{center}
If we measure $0$, then we start all over again. If we rather measure $2$, then we braid anyons $4$ and $5$ and we remeasure, a process we do until we measure a $0$ and are back to the initial configuration of input and ancilla. Alternatively, as suggested by Parsa Bonderson, we measure anyons $4$ and $5$ and hope to measure $0$ which would achieve the desired fusion. If we measure $2$ instead, we are simply back to the initial wrong outcome and proceed just like before. In summary, we could see the algorithm as an alternation of measuring anyons $4$ and $5$ on one hand and measuring anyons $5$, $6$, $7$, $8$ on the other hand, until the expected fusion is achieved. Anyhow, after possibly several tries, we have now fused anyons $4$ and $5$. We will continue fusing the ancilla into the input one step further by now measuring anyons $3$ and $4$, like below.
\begin{center}
\epsfig{file=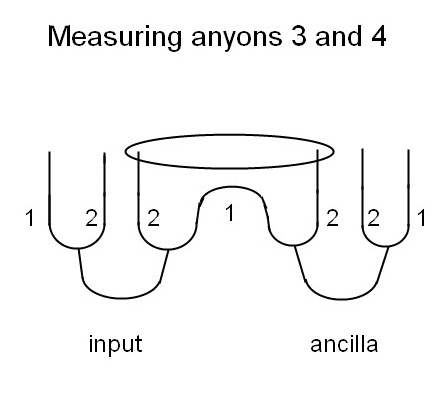, height=5.5cm}
\end{center}
The measurement outcome is either $0$, $2$ or $4$.
\begin{itemize}
\item
If the measurement outcome is $0$, then we obtain the proposed gate.
\item
If the measurement outcome is $4$, we do two additional steps: we fuse anyons $3$ and $4$, and we then fuse anyons $3$ and $4$ again where it is understood that the anyons get renumbered from left to right after each action. For both fusions, the outcome is unique (4 for the first fusion, 2 for the second fusion) by the theta rule and by the fusion rules respectively. It yields the gate
$$\begin{array}{ccc}\begin{pmatrix}
b&0\\
0&a
\end{pmatrix}&\text{instead of the gate}&\begin{pmatrix} a&0\\0&b\end{pmatrix}\end{array}$$
Since all our gates are defined up to some overall complex scalar, this is simply the inverse gate. Then, by doing a random walk, we are able to eventually produce the gate we want. No need of recovery here: only an iteration of the overall process.
\item If the measurement outcome is $2$, suppressing the bubble with three external edges introduces some minus signs when the input edge and the ancilla edge carry a distinct label. However, these minus signs get exactly compensated after doing an F-move on the edge of the measurement. Then, by running an interferometric measurement on the last three anyons, we get back to the initial configuration after the first fusion of the ancilla into the input with the only difference that the middle horizontal edge could now carry the label $3$ instead of $1$, depending on the outcome of the interferometric measurement. However, whatever be the label, the recovery is complete and we start all over again by measuring anyons $3$ and $4$.

\end{itemize}
Below are some figures which illustrate the different measurement outcomes and their treatment in the same order in which they were discussed above.\\
\vspace{-0.4cm}
\epsfig{file=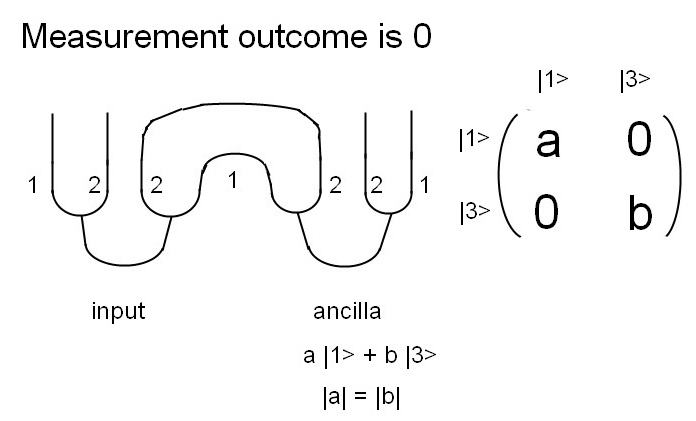, height=7cm}
\begin{center}
\epsfig{file=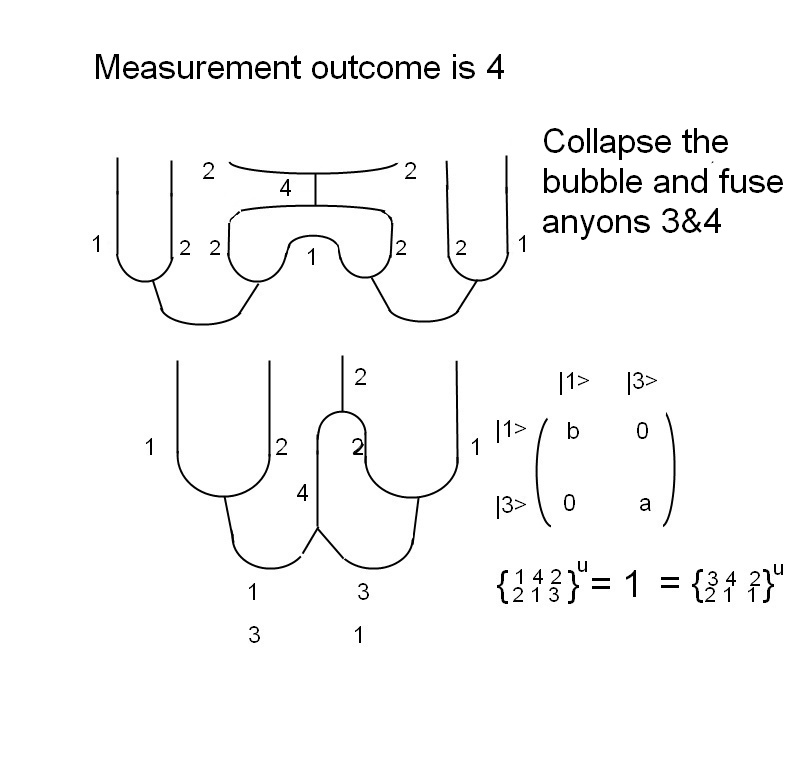, height=10cm}\end{center}
\vspace{-0.9cm}
\epsfig{file=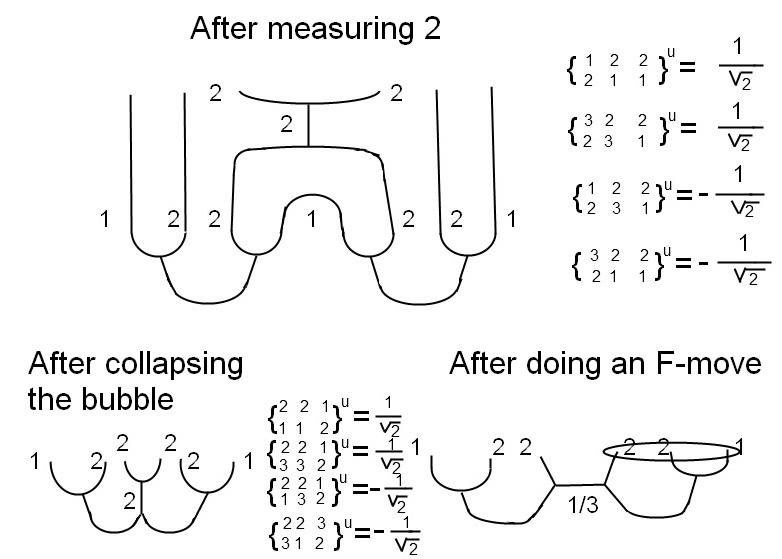, height=8.7cm}\\\\
In the second series of unitary $6j$-symbols given above, if the upper right corner number is $3$ instead of $1$, get the same values with opposite signs.

\section{The qutrit $2222$}
Before we state the theorem, we will outline how we fuse the "last" anyon from the input and the "first" anyon from the ancilla with a "forced" measurement to $0$. We measure the collective charge of both anyons. If the outcome is $0$, success. If the outcome is $2$, by forthcoming Eq. $(1.1)$ we see that after doing an F-move, we have the following configuration
\begin{center}
\epsfig{file=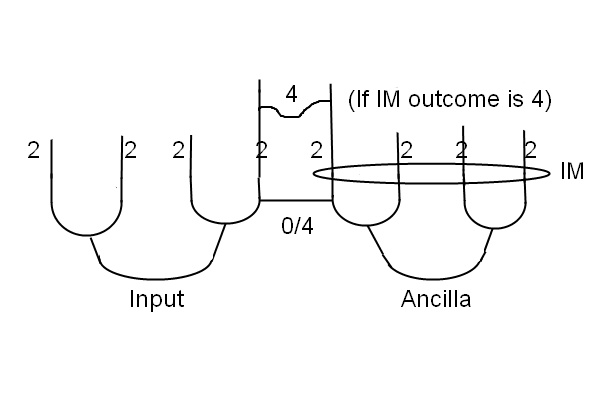, height=7cm}
\end{center}
Like on the figure above, run an interferometric measurement. If we measure $0$, we are back to the initial configuration and try again. If we measure a $4$, it will suffice to fuse a pair of $4$'s like on the figure in order to cancel the two $4$ charge lines. This is described as a Freedman fusion operation in \cite{CL}, but it can also be interpreted as
\begin{center}
\epsfig{file=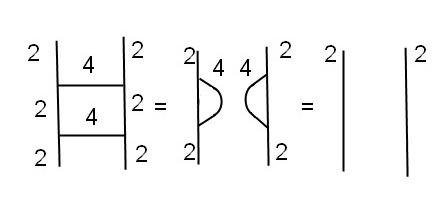, height=5cm}
\end{center}
after doing an F-move on the $0$ labeled edge between the two $4$ charge lines.
Finally if the measurement outcome is $4$, then use the trick from before and fuse a pair of $4$'s like on the figure.
\begin{center}
\epsfig{file=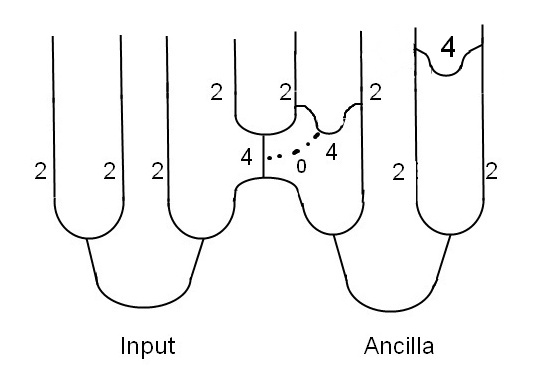, height=9cm}
\end{center}
By a careful inspection at the unitary $6j$-symbols,
 $$\begin{array}{ccc}\left\lbrace\begin{array}{ccc}2&2&2\\2&2&4\end{array}\right\rbrace^u=\,-\,\frac{1}{\sqrt{2}}&\&&
 \left\lbrace\begin{array}{ccc}2&2&0\\2&2&4\end{array}\right\rbrace^u=\left\lbrace\begin{array}{ccc}2&2&4\\2&2&4\end{array}\right\rbrace^u=\frac{1}{2}\end{array},$$
 we have accomplished the fusion we wanted, except the $|2>$ trit from the ancilla has been switched to $-|2>$. It then suffices to fuse another pair of $4$'s like on the figure in order to switch the $-|2>$ trit of the ancilla back to $|2>$ by the following Lemma.

\newtheorem{Lemma}{Lemma}
\begin{Lemma} \hfill\\
\begin{center}
\epsfig{file=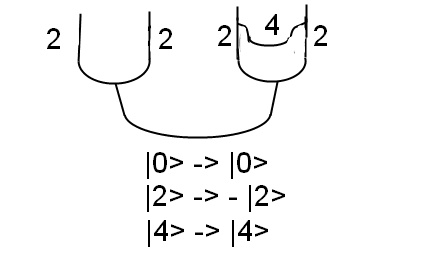, height=4cm}
\end{center}
\end{Lemma}

We now state the Theorem.
\newtheorem{Theorem}{Theorem}
\begin{Theorem}
Given an ancilla $|0> + \sqrt{2}\,e^{i\al_2}|2>+|4>$, the following protocol
\begin{center}
\epsfig{file=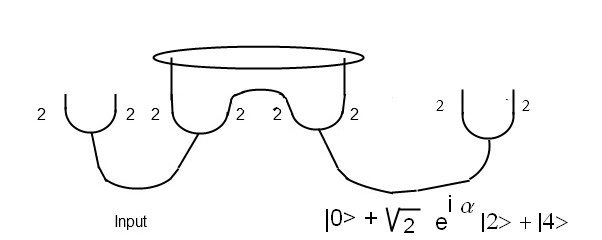, height=5cm}
\end{center}
There are three cases. \\

\textit{Case $(i)$}. The outcome of the measurement is $0$. Then, we have formed the gate
$$\begin{pmatrix}
    1&0&0\\
    0&e^{i\,\al}&0\\
    0&0&1\end{pmatrix}$$

\textit{Case $(ii)$}. The outcome of the measurement is $4$. Then, fuse anyons $3$ and $4$ necessarily to $4$. Then fuse anyons $3$ and $4$ necessarily to $2$. We have formed the same gate as in $(i)$.\\

\textit{Case $(iii)$}. The outcome of the measurement is $2$. Then, fuse anyons $3$ and $4$ necessarily to $2$. Then fuse anyons $3$ and $4$ necessarily to $2$. We have then formed the inverse gate
$$\begin{pmatrix}
e^{i\,\al}&0&0\\
0&1&0\\
0&0&e^{i\,\al}\end{pmatrix}$$
To finish do a random walk in order to eventually produce the gate of $(i)$.
\end{Theorem}

\textsc{Proof.} When the outcome is zero, the bubble removal introduces a factor $\frac{1}{\sqrt{2}}$ in front of the $|2>$ trit, which compensates exactly the $\sqrt{2}$ from the ancilla. We get the phase gate announced. \\

We next deal with Case $(ii)$. On the figure below, all the unitary $6j$-symbols that are useful to the computations and got omitted take the value $1$. This case illustrates why it is important to have an ancilla that has identical phases in $|0>$ and $|4>$. We obtain the same gate as in $(i)$.
\begin{center}
\epsfig{file=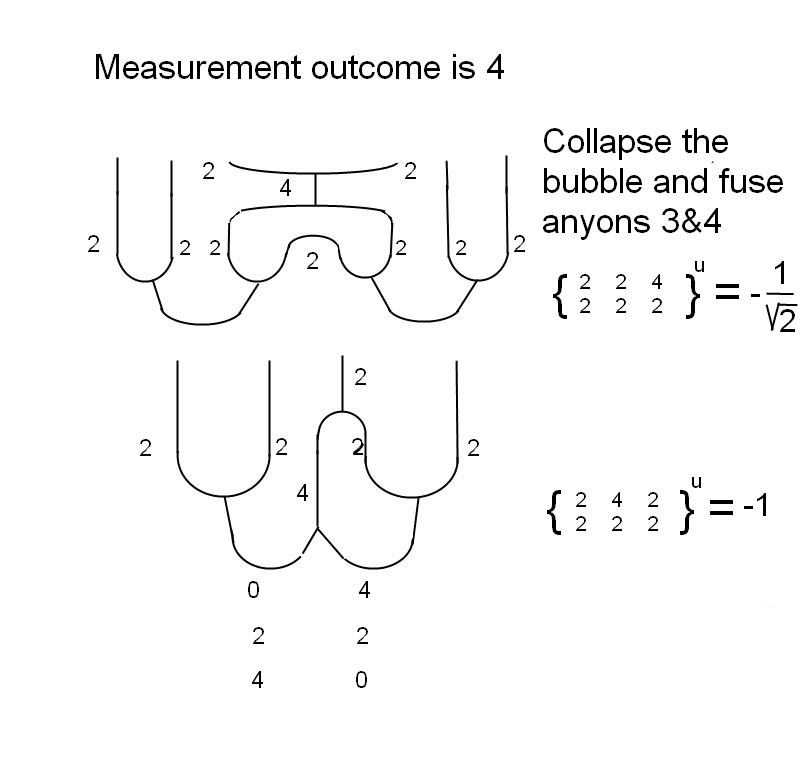, height=10cm}
\end{center}
\indent Finally, when the measurement outcome is $2$, the $(|2>,|2>)$ contribution vanishes as specific to this theory, we have
\begin{equation}\left\lbrace\begin{array}{ccc}2&2&2\\2&2&2\end{array}\right\rbrace^u=0,\end{equation}
We obtain the $2$-qutrit vector
$$|2>\otimes(|0>-|4>)+e^{i\al}\,(|0>-|4>)\otimes |2>$$
Next we fuse anyons $3$ and $4$ and after that we fuse anyons $3$ and $4$ again (same protocol as when the measurement outcome was $4$). \\
We claim that the outcome from the second fusion must be $2$ and not $0$ or $4$. This is partly because the $(|0>-|4>)\otimes |2>$ contribution of the superposition would vanish if the fusion outcome were $0$ or $4$ by the fusion rules and partly because
the $|2>\otimes\, (|0>-|4>)$ contribution of the superposition would also vanish if the fusion outcome were $0$ or $4$ since all the $6j$-symbols that are involved are equal. Thus, there wouldn't be any contribution left. And so, we will never measure $0$ or $4$ as illustrated on the figure below.

\noindent The straightforward computations lead to producing the gate
$$\begin{pmatrix}
e^{i\,\al}&0&0\\
0&1&0\\
0&0&e^{i\,\al}\end{pmatrix}$$
which, up to overall phase $e^{i\al}$ is the inverse gate of the one from the first two cases, that is
 $$\begin{pmatrix}
1&0&0\\
0&e^{-i\al}&0\\
0&0&1\end{pmatrix}$$

\noindent In order to always produce the gate of $(i)$ or $(ii)$, it will suffice to iterate the process, that is do a random walk until we obtain the desired gate.
\begin{center}
\epsfig{file=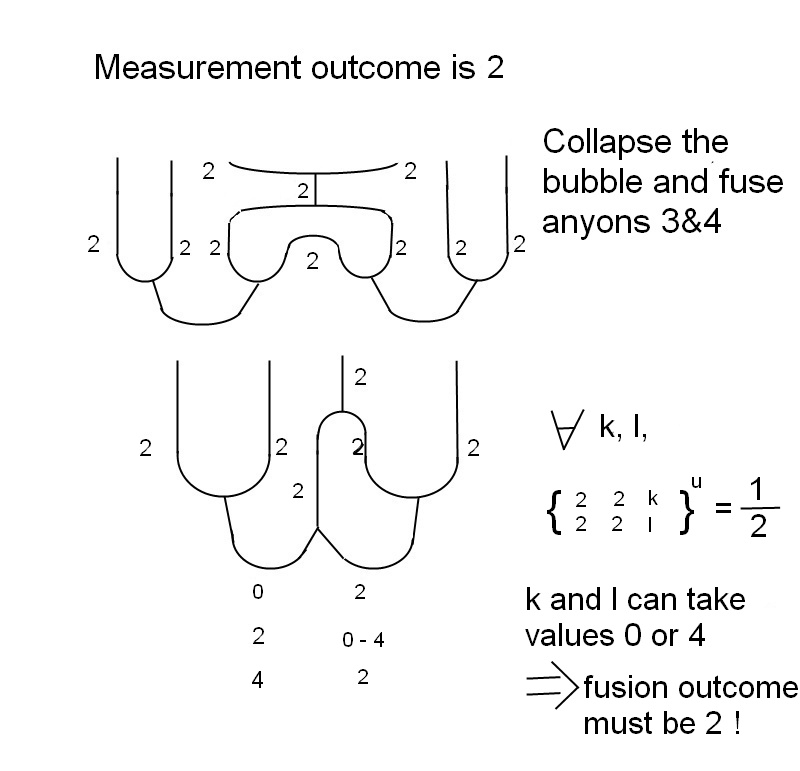, height=10cm}
\end{center}

\noindent\textbf{Acknowledgements.}
The author thanks Michael Freedman for quite helpful and inspiring discussions and Parsa Bonderson for many nice comments. She thanks Matt Hastings for enlightening discussions.\newpage
\chapter{Irrational phase gates}
\begin{center}
Note
\end{center}
The author thanks Michael Freedman for providing these thrilling problems, for guiding her with endless generosity in time and encouragements and for communicating his own excitement during the pursuit of this research.
\begin{center}
Abstract
\end{center}
In Chapter $1$, it is shown with respect to two different anyonic systems, the same as those considered in this paper, how to fuse an ancilla into the input in order to form a gate. In each case, the ancilla must satisfy to some adequate amplitude ratios properties. In the present chapter, we show how to prepare such ancillas, leading to infinite order phase gates. We obtain universal $1$-qubit and $1$-qutrit gates. Together with \cite{CL2} and
 \cite{CL4} in which are respectively shown how to make a $2$-qubit and a $2$-qutrit entangling gate, we get qubit and qutrit gate sets that are universal for quantum computation.

\section{Introduction}
The framework of this work is the Kauffman-Jones version of $SU(2)$ Chern-Simons theory at level $4$.
By braiding only of $4$ anyons of topological charges $1221$ or $2222$, we obtain gates that generate a finite subgroup of $SU(2)$ or $SU(3)$ respectively. By adding some fusion operations, it is possible to enlarge the size of the respective groups but it is still not possible to make them infinite and dense. The case of the qutrit was extensively studied in \cite{BL} and \cite{CL}. Measurements assisted gates appear to become a necessity for universality. In \cite{F}, a general protocol for $1$-qubit and $1$-qutrit "gate hunting" is provided which uses braids and interferometric measurement. It is composed of a few main steps which are summarized on the figure below. The idea is to fuse the right most anyon from the input and the left most anyon from the ancilla by doing a measurement with recovery which leaves a pair of anyons in position $4-5$, then do braids involving all the other anyons, then do an interferometric measurement forced to $0$ on anyons $5,6,7,8$ to separate the input and the ancilla. The latter measurement having an easy recovery if the outcome is not $0$. There are three major difficulties in this protocol. First, we must check for no-leakage of quantum information when we operate the interferometric measurement. This leads to conditions on the ancilla we use. These conditions depend heavily on the braids we perform before running the interferometric measurement. Next, we must be able to prepare an ancilla which pleases these conditions. This is not always easy. For the qubit $1221$, some ancillas not issued from braids are provided in \cite{QIP} which are useful in this general protocol. Third, after the interferometric measurement, the input and the ancilla will be topologically separated but could be entangled. It can be very hard to find a disentangler which uses braids and perhaps also measurements, though it was shown possible in some cases. For instance the $2$-qutrit entangler of \cite{CL2} is used in \cite{F} as a disentangler. We may have to run more interferometric measurements and have to check for no-leakage again. Last, after overcoming all these difficulties, we must hope for an interesting gate, that is a gate which is not already in the image of the braid group.
\begin{center}
\epsfig{file=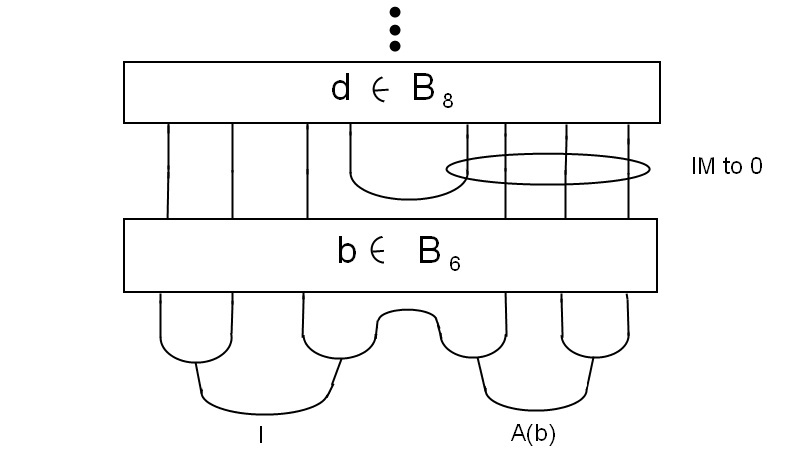, height=7cm}
\end{center}
On the figure above, the letter $I$ stands for "input", the letter $A$ for "ancilla", the letter $b$ for "braid", the letter $d$ for "disentangler".\\
This theoretically interesting protocol did not so far yield any interesting results. In the present paper, we show how to produce infinite order phase gates for both the qubit and the qutrit by using a different protocol.\\
The paper is divided into two parts, one that deals with the qubit $1221$ and one that deals with the qutrit $2222$. The scheme is the same for the $1$-qubit gate or the $1$-qutrit gate. First, we do an ancilla preparation. Second we fuse the ancilla into the input following Chapter $1$ in order to form the gate.
In Chapter $1$ it is shown that an ancilla
$$x|1>+y|3>\;\text{with $|x|=|y|$}$$ for the qubit $1221$ can be turned into a gate $$\begin{pmatrix}x&0\\0&y\end{pmatrix}$$
And an ancilla $$|0>+\,\sqrt{2}\,e^{i\al}|2>+|4>$$ for the qutrit $2222$ can be turned into a gate $$\begin{pmatrix}1&&\\&e^{i\al}&\\&&1\end{pmatrix}$$
The present paper focuses on ancilla preparations. \\
The ancilla preparation for the qubit can be divided into two parts. First, we must find a way to create an interesting relative phase, say $\theta$. Second, we must have equal complex norms in $|1>$ and $|3>$ in order to avoid leakage during the fusion of the ancilla into the input. An idea to get an interesting phase is simply to add two complex numbers. An idea to get equal norms is to notice that for two given complex numbers $a$ and $b$, we have
$$|a\bar{b}|=|b\bar{a}|$$
Thus, suppose we can prepare an ancilla $\left(\begin{array}{l}a\\b\end{array}\right)$ with interesting relative phase and suppose we are able to prepare the conjugate ancilla $\left(\begin{array}{l}\bar{a}\\\bar{b}\end{array}\right)$. By the following Freedman operation
\begin{center}
\epsfig{file=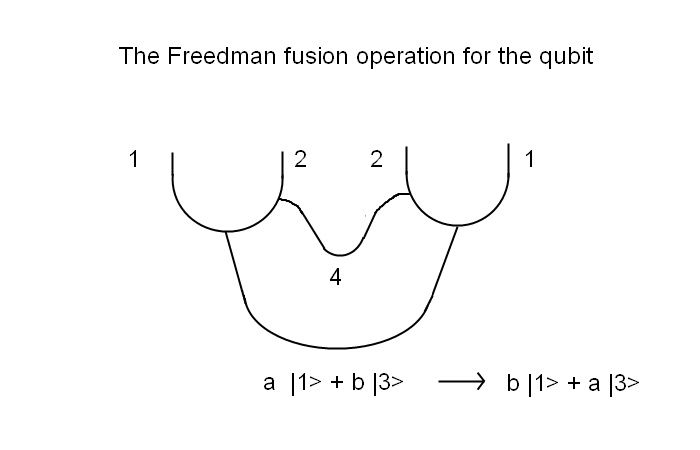, height=6cm}
\end{center}
we also know how to prepare the ancilla $\left(\begin{array}{l}\bar{b}\\\bar{a}\end{array}\right)$.
Then, we can fuse the two ancillas together in order to obtain a third ancilla which is
$$\left(\begin{array}{l}a\,\bar{b}\\b\,\bar{a}\end{array}\right)$$
As a matter of fact, the complex norms in $|1>$ and $|3>$ are equal and the relative phase is now $2\theta$ instead of $\theta$, which is still an "interesting" relative phase. The goal is then reached.\\

For the qutrit $2222$, it is slightly more difficult to prepare an adequate ancilla. While the idea to get an interesting phase between the $|0>$ trit and the $|2>$ trit or between the $|2>$ trit and the $|4>$ trit is rather similar as for the qubit - that is use any braids, any measurements and any anyons to make such a qutrit - there are two main difficulties once this is achieved.

First, we must have equal phases in $|0>$ and $|4>$, else the "ancilla turning into gate" protocol described in lengthy details in Chapter $1$ will not work.

Second, we must have the correct norm ratios like provided in the discussion above. Again, the strategy we used for the qubit won't apply here and we must come up with new ways.

In order to solve the problem of "identical phases" in $|0>$ and $|4>$, it will suffice to use a process of
"pair fusion" which allows by fusing a pair of $2$'s into anyons $2$ and $3$ of the qutrit to transfer and clone the contribution from the $|2>$ trit onto the $|0>$ and $|4>$ trits (then get identical phases, like aimed at), while the two contributions in $|0>$ and $|4>$ get added to yield the contribution in $|2>$.

The most tricky part remains to have the correct norm ratios. In order to achieve this, we use a subtle combination of three protocols, namely the pair fusion that was just mentioned, together with what we called the "qutrit fusion" and the "qutrit projection".
The qutrit projection allows, given a qutrit ancilla, to eliminate exactly one of the two contributions arising from the $|0>$ trit or the $|4>$ trit.
What is important is that we are left with two trits which carry the interesting relative phase. If furthermore, the ratio of the norms in $|0>$ (or $|4>$) and $|2>$ equals $\sqrt{2}$, then by applying a pair fusion, we will get the three desired things: adequate norms ratios, the same interesting relative phase between the $|0>$ and $|2>$ trits and identical phases for the $|0>$ and $|4>$ trits. Note where the importance of the trit elimination precisely is. If you don't do it, you can still get the same phases in $|0>$ and $|4>$ by pair fusion and perhaps still have an interesting relative phase, but you will loose the adequate norms ratios because you will have added two distinct complex numbers during the pair fusion process, which can create a new uncontrolled amplitude. Let us conclude by saying a word about the qutrit fusion. Suppose we are able to make an ancilla $a|0>+b|2>$ with an interesting relative phase and its conjugate ancilla.
Given these two ancillas, the qutrit fusion allows by fusing them to get a third ancilla like below.
$$\begin{array}{cc}\begin{array}{l}|0>\\|2>\\|4>\end{array}&\begin{pmatrix}a\\b\\0\end{pmatrix}\end{array}, \;
\begin{array}{cc}\begin{array}{l}|0>\\|2>\\|4>\end{array}&\begin{pmatrix}\bar{a}\\\bar{b}\\0\end{pmatrix}\end{array}
\longrightarrow\begin{array}{cc}\begin{array}{l}|0>\\|2>\\|4>\end{array}&\begin{pmatrix}a\bar{b}\\\frac{1}{\sqrt{2}}b\bar{a}\\0\end{pmatrix}\end{array}$$
The latter resulting ancilla has the appropriate norms ratio in order to successfully benefit from the pair fusion operation and twice the interesting relative phase we started from.
It is an ancilla which is ready to be turned into a good final ancilla by pair fusion like in our discussion above.
While, as we just saw, the qutrit pair fusion's beneficial outcome relies on the nice shape of the ancilla issued from the qutrit fusion protocol, it is very important to note that the qutrit fusion's beneficial outcome is itself made possible by the prior qutrit projection. This fact will be enlightened later on when we get into the details of the qutrit fusion. Thus, in this protocol, all the different parts fit together and intertwine with each other beautifully. \\\\We illustrate the latter comment by outlining the different steps.
Given an ancilla with interesting relative phase in $|0>$ (or $|4>$) and $|2>$ and its conjugate ancilla,
\begin{itemize}
\item The \textbf{qutrit projections} allow a successful qutrit fusion of both ancillas.\\
\item The \textbf{qutrit fusion} allows adequate norms ratios for a successful qutrit pair fusion.\\
\item The \textbf{qutrit pair fusion} allows identical phases in $|0>$ and $|4>$ and correct norms ratios in preparation for the ancilla fusion into the input of Chapter $1$.
\end{itemize}
In light of the outlines provided above, we are ready to now describe the ancilla preparations themselves. First for the qubit $1221$, then for the qutrit $2222$. \\The protocols which produce an ancilla ready to be turned into an irrational phase gate appear on pages $31$ and $39$ for the qubit and qutrit respectively.
\section{Irrational qubit phase gate}
\subsection{First ancilla $A_1$}
The starting point is four anyons of topological charge $1$ forming the qubit $|2>$, together with a pair of $2$'s out of the vacuum. We do two braids, followed by two fusion measurements and one braid. The protocol is represented on the figure below.
\begin{center}
\epsfig{file=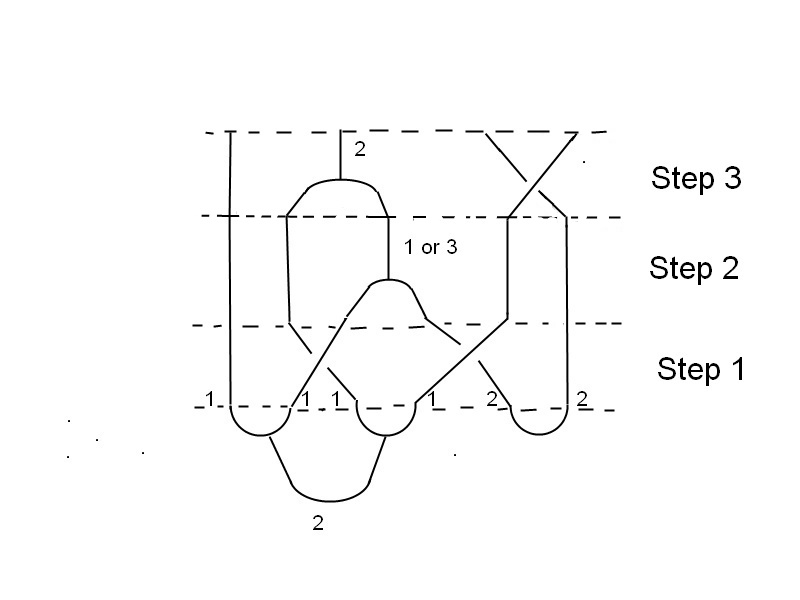, height=10cm}
\end{center}

Below, we will explain the ideas that are hidden behind the three steps of the figure. \\\\
\textit{Step $1$}. The left braid creates a superposition of $|0>$ and $|2>$ for the qubit $1111$.
The second braid creates a superposition of $|1>$ and $|3>$ when braiding $(1122)_0$. The latter charge line will in the end carry the quantum information for the ancilla.
\begin{center}
\epsfig{file=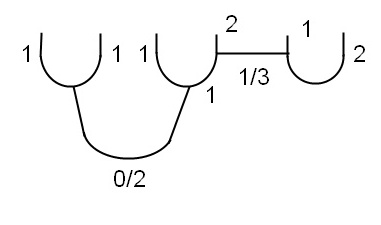, height=4cm}
\end{center}
\textit{Step $2$}.
This fusion and the subsequent one are really at the origin of the interesting relative phase. The idea is based on the fact that whether the measurement outcome of the fusion is $1$ or $3$, for exactly one of the qubits $|1>$ or $|3>$ from the right horizontal charge line, both contributions of $|0>$ and $|2>$ remain on the left horizontal charge line. This is a consequence of the fusion rules. Two anyons of respective topological charges $3$ and $0$ cannot fuse into an anyon of topological charge $1$ and two anyons of respective topological charges $1$ and $0$ cannot fuse into an anyon of topological charge $3$. From now on, suppose the fusion measurement outcome is $1$, like illustrated below.
\begin{center}
\epsfig{file=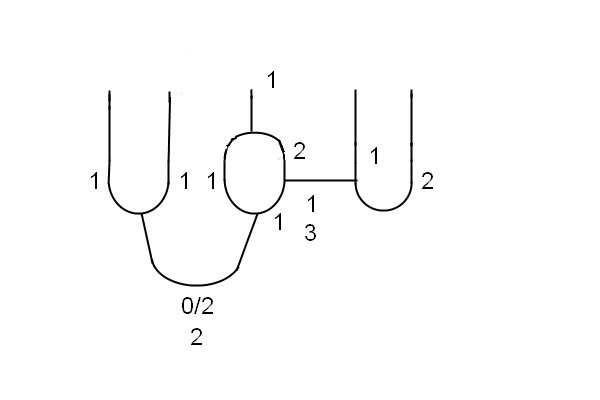, height=7cm}
\end{center}
\textit{Step $3$}. When the second fusion occurs (and we force the outcome from this fusion to be $2$), for the $1$ horizontal charge line, the bubble suppression results in summing two contributions: one from the $0$ horizontal charge line and one from the $2$ horizontal charge line. This sum yields a quite interesting phase. Whereas for the $3$ horizontal charge line, there is only one term, namely the one from the $2$ horizontal charge line. Now we get a quite interesting \textit{\textbf{relative}} phase. It remains to braid the last two anyons in order to get to the qubit shape $1221$.
\begin{center}
\epsfig{file=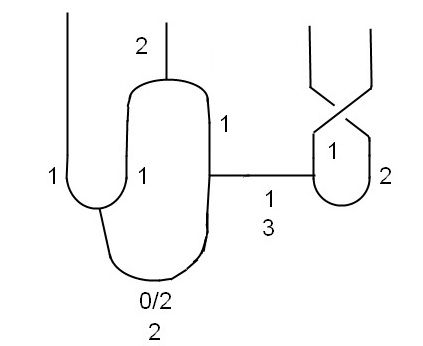, height=5cm}
\end{center}

\noindent We now go over the computations themselves. First, recall the matrix for a $\sigma_2$-braid on four anyons of topological charge $1$.\\\\

$$G_2[1,1,1,1]=\begin{pmatrix} \frac{e^{\frac{i\pi}{4}}}{\sqrt{3}}&&\sqrt{\frac{2}{3}}\,e^{-i\frac{5\pi}{12}}\\&&\\
\sqrt{\frac{2}{3}}\,e^{-i\frac{5\pi}{12}}&&\frac{e^{-i\frac{\pi}{12}}}{\sqrt{3}}\end{pmatrix}$$
\begin{center}
\epsfig{file=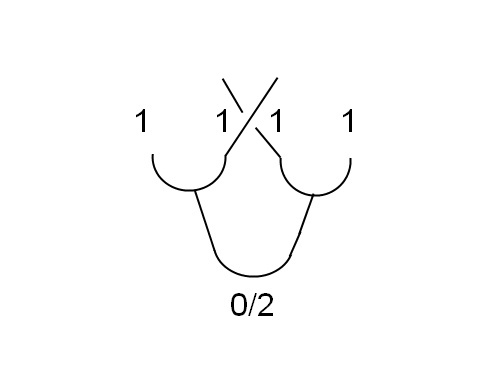, height=6cm}
\end{center}

\noindent We also recall from \cite{QIP} the matrix for the action by a $\sigma_2$-braid on four anyons $1122$.
\vspace{-0.65cm}
\begin{center}
\epsfig{file=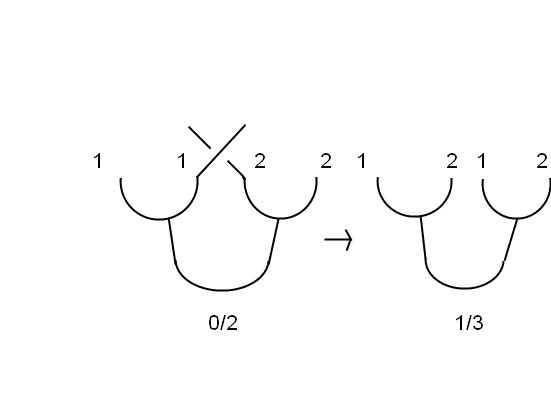, height=8cm}
\end{center}
$$\begin{array}{l}\qquad\qquad\qquad\qquad\qquad\qquad |0>\qquad\qquad |2>\\\\G_2[1,1,2,2]=\begin{array}{cc}\begin{array}{l}|1>\\\\|3>\end{array}&\begin{pmatrix} \frac{1}{\sqrt{2}}\,e^{i\frac{2\pi}{3}}&&\frac{1}{\sqrt{2}}\\&&\\\frac{1}{\sqrt{2}}\,e^{-i\frac{5\pi}{6}}&&-\frac{i}
{\sqrt{2}}
\end{pmatrix}\end{array}\end{array}$$
$$\begin{array}{l}\\\end{array}$$
So, after the two braids of Step $1$, we read from the first column of $G_2[1,1,2,2]$ and the second column of $G_2[1,1,1,1]$ that we have the superposition \\
$\begin{array}{l}\\\end{array}$$
$$\begin{array}{cc}\begin{array}{l}|01>\\\\|03>\\\\|21>\\\\|23>\end{array}&\begin{pmatrix}\frac{1}{\sqrt{3}}
\,e^{i\frac{\pi}{4}}\\
\\\star\\\\\frac{1}{\sqrt{6}}\,e^{i\frac{7\pi}{12}}\\\\\frac{1}{\sqrt{6}}
\,e^{-i\frac{11\pi}{12}}
\end{pmatrix}\end{array}
$ $\begin{array}{l}\\\epsfig{file=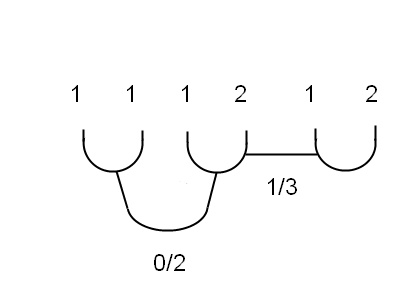, height=4cm}\end{array}$
$$\begin{array}{l}\\\end{array}$$
Note, we only provided the useful contributions. Indeed, a triangle inside a net can be removed at the cost of a multiplication by a scalar, like on the figure below.\\
\epsfig{file=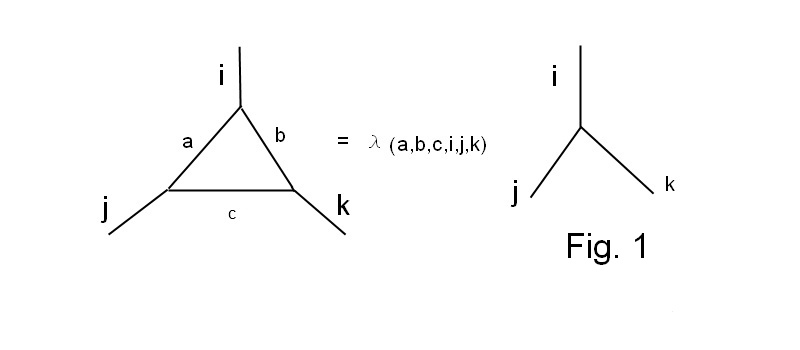, height=6.5cm}\\
When we fuse anyons $3$ and $4$ and measure a $1$,
we see with the fusion rules that only the three $2$-qubits $|01>$, $|21>$, $|23>$ contribute to the measurement.
After the triangle collapse, we get a superposition\\

$$\begin{array}{cc}\begin{array}{l}|01>\\\\|21>\\\\|23>\end{array}&\begin{pmatrix}
2\sqrt{2}\,e^{i\frac{\pi}{4}}\\\\e^{-i\frac{5\pi}{12}}\\\\\sqrt{3}\,e^{-i\frac{11\pi}{12}}
\end{pmatrix}\end{array}$$
$\begin{array}{l}\\\end{array}$\\
After the second fusion, we get an ancilla qubit $1212$. The contributions are as follows, where the brackets denote the unitary $6j$-symbols.
\begin{itemize}
\item For the $|1>$ qubit:
\begin{equation}
\left\lbrace\begin{array}{ccc} 1&1&2\\1&1&0\end{array}\right\rbrace^{u}\;2\;\sqrt{2}\;e^{\frac{i\pi}{4}}\;+\;
\left\lbrace\begin{array}{ccc} 1&1&2\\1&1&2\end{array}\right\rbrace^{u}\,e^{-\frac{i5\pi}{12}}=\sqrt{7}\,ArcTan\left(\frac{14+3\sqrt{3}}{13}\right)
\end{equation}

\item For the $|3>$ qubit:
\begin{equation*}
\left\lbrace\begin{array}{ccc} 1&1&2\\1&3&2\end{array}\right\rbrace^{u}\,\sqrt{3}e^{-\frac{i11\pi}{12}}=\sqrt{3}e^{-\frac{i11\pi}{12}}
\end{equation*}
\end{itemize}

Then, after R-move, we finally get an ancilla qubit $1221$, which we name $A_1$. We have
$$A_1=\frac{1}{\sqrt{10}}(\sqrt{7}e^{i\,arctan(\frac{-14-5\sqrt{3}}{11})}|1>+\sqrt{3}\,e^{-i\frac{\pi}{12}}|3>)$$
Note, if we apply a $\sigma_2$-full twist on this ancilla, we get another ancilla:
$$A^{'}_1=\frac{1}{\sqrt{10}}(|1> + 3\,|3>)$$
More generally, by doing more braids, we can make many interesting ancillas for the qubit $1221$ which we would not be able to make by braiding only.
\newtheorem{Remark}{Remark}
\begin{Remark}
If the first fusion outcome is $3$ instead of $1$, it can be read from the diagrams and from the values of the unitary $6j$-symbols that we get an ancilla where the contributions in $|1>$ and $|3>$ are swapped. Hence, it does not matter whether we measure a $1$ or a $3$ after the first fusion. Note also that one ancilla can be obtained from the other one by a Freedman type fusion operation mentioned in the introduction. After the second fusion however, we want to measure a $2$. If we were to measure a $0$ instead, we would start the ancilla preparation all over again.
\end{Remark}
\subsection{Conjugate ancilla $\overline{A_1}$}
This section enlightens quite particularly the necessity of the protocol that produced the first ancilla. Indeed, a quite specific and interesting fact about the matrix $G_2[1,1,1,1]$ is that its two anti-diagonal coefficients are identical.
So that the conjugate transpose of this matrix equals its conjugate, that is the inverse matrix of $G_2[1,1,1,1]$ is the conjugate matrix of $G_2[1,1,1,1]$. Thus, if
$$\left(\begin{array}{l}x_0\\x_2\end{array}\right)$$ denotes the second column of the matrix $G_2[1,1,1,1]$, the second column of $G_2[1,1,1,1]^{-1}$ is simply
$$\left(\begin{array}{l}\overline{x_0}\\\overline{x_2}\end{array}\right)$$
After a quick glance at the first column of the matrix $G_2[1,1,2,2]$ and at the R-matrix
$$R[1,2]=\begin{pmatrix} -e^{i\frac{\pi}{3}}&0\\0&-e^{-i\frac{\pi}{6}}\end{pmatrix},$$
we see that the two right most braids of the protocol of $\S\,2.1$ cancel exactly. This could also be seen as a Reidemeister's move.
Since removing bubbles from the fusion measurements only introduces real numbers in the computations, it appears clearly that if we do the exact same protocol as before with the exception of an inverse $\sigma_2$-braid on $(1111)_2$ instead of a $\sigma_2$-braid, we will produce the conjugate ancilla $\overline{A_1}$. For clarity, we provide below the protocol which produces $\overline{A_1}$ and which simply differs from the previous protocol by one braid being replaced by its inverse braid.
\begin{center}
\epsfig{file=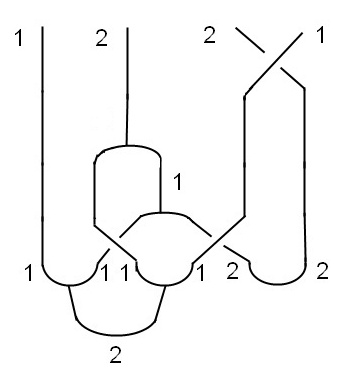, height=6cm}
\end{center}

\noindent In the next section, we explain how by fusing $A_1$ and $\overline{A_1}$ together, we succeed to make an ancilla which has equal norms of $|1>$ and $|3>$ and where the interesting relative phase from $A_1$ is doubled.

\subsection{Final ancilla $A_f$}

The general protocol described here applies to the qubit $1221$.
It takes two ancillas
$$\begin{array}{ccc}\left(\begin{array}{l}a\\b\end{array}\right)&\text{and}&\left(\begin{array}{l}x\\y\end{array}\right)
\end{array}$$ and returns a third ancilla which is
$$\begin{array}{cccc}\text{either}&\left(\begin{array}{l}ax\\by\end{array}\right)&\text{or}&\left(\begin{array}{l}ay\\bx\end{array}\right)
\end{array}$$

\noindent We take two such ancillas. Following ideas from \cite{F}Chapter $1$, we measure anyons $4$ and $5$. If the outcome is $0$, we have fused the two anyons. If the outcome is $2$, we run an interferometry on anyons $5,6,7,8$ with a desired outcome of $0$. Then we measure anyons $4$ and $5$ again and iterate the process. If the interferometric measurement outcome is $2$, we measure again anyons $4$ and $5$ and hope to measure $0$. If we measure $2$ , we run an interferometric measurement on anyons $5,6,7,8$ again and proceed like before depending on the outcome. We remove the pair of $1$'s. \\
\underline{Alternatively}, since we are dealing with ancilla preparations, we fuse anyons $4$ and $5$ together. If the outcome of the fusion is $0$, fine. If the oucome of the fusion is $2$, we start all over again with two fresh ancillas. In this alternate procedure, there is no pair of $1$'s to remove at the end.\\
We get the well-known picture of \cite{F}Chapter $1$.
\begin{center}
\epsfig{file=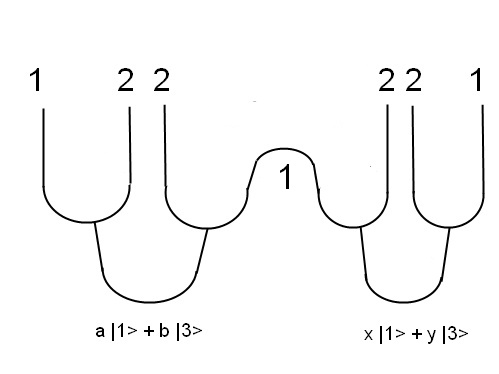, height=6cm}
\end{center}
The idea next is to fuse anyons $3$ and $4$. We somehow got inspired by \cite{BK}.
\begin{center}
\epsfig{file=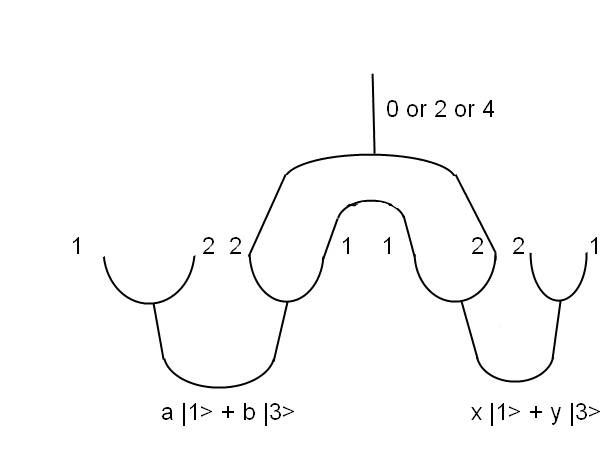, height=7cm}
\end{center}

Our initial state is the tensor product $(a|1> +b|3>)\otimes (x|1>+y|3>)$.
\begin{itemize}
\item If we measure a $0$, since the quantum dimensions of $1$ and $3$ are equal,
    we get the ancilla
    $$a\,x\,|1>\, + \,b\,y\,|3>$$
    where the contribution in $|1>$ comes from the tensor $|1>\otimes |1>$ and the contribution in $|3>$ comes from the tensor $|3>\otimes |3>$.
\item If we measure a $4$, then do a second fusion, namely of anyons $3$ and $4$, whose outcome is necessarily a $2$. We get the ancilla
    $$a\,y\,|1>\,+\,b\,x\,|3>,$$
    where the contribution in $|1>$ comes from the tensor $|1>\otimes |3>$ and the one in $|3>$ from the tensor $|3>\otimes\, |1>$.
\item If we measure a $2$, we fail, but we may be able to retrieve one of the two ancillas by doing an extra fusion measurement. Namely, we will retrieve the ancilla to the left $a|1> + b|3>$ if anyons $3$ and $4$ fuse into $2$ and we will retrieve the ancilla to the right $x|1> + y|3>$ if anyons $2$ and $3$ fuse into $2$. Of course in either case, there is no insurance that the fusion outcome will be $2$.
\end{itemize}
\noindent\underline{Alternatively}, we could measure anyons $3$ and $4$ instead of fusing them. If the outcome is $0$, no change. If the outcome is $4$, we can fuse anyons $3$ and $4$ and then conclude like before. If the outcome is $2$, the nice thing is that we are able to retrieve both ancillas instead of at most one. The process is the following. Run an interferometric measurement on the last three anyons whose oucome can be either $1$ or $3$. If the outcome is $3$, then fuse a pair of $4$'s into anyons $3$ and $4$ in order to simulate having measured a $1$. Next, bring a pair of $1$'s in between anyons $3$ and $4$ and measure the last four anyons by interferometry. If measure $0$, end of the process. If rather measure $2$, then braid anyons $4$ and $5$ and remeasure by interferometry anyons $5,6,7,8$. Iterate in the case when the outcome of the latter measurement is $2$. \\
By $\S\,2.1$, $\S\,2.2$ and $\S\,2.3$, we are able to make the ancilla
$$\boxed{\begin{array}{l}A_f=\,|1>\,+\,e^{i(2\theta^{'}\,+\,\frac{\pi}{6})}\,|3>\\\\\text{with}\;\;
\theta^{'}=\,ArcTan(\frac{-14-5\sqrt{3}}{11})\end{array}}$$
Set $$\theta=2\theta^{'}\,+\,\frac{\pi}{6}$$
Then, by the protocol of Chapter $1$, we are able to make the gate
$$\Lambda(e^{i\theta})=\begin{pmatrix}1&0\\0&e^{i\theta}\end{pmatrix}$$
or the inverse gate $$\Lambda(e^{-i\theta})=\begin{pmatrix}1&0\\0&e^{-i\theta}\end{pmatrix}$$
By doing a random walk, we are able to make the gate $\Lambda(e^{i\theta})$, see for instance \cite{BK}.
We conclude by using a statement of \cite{OL} which we recall below.
\begin{Theorem}(by Olmsted) If $x$ is rational in degrees, then the only possible rational values of the tangent are: $$tan\,x=0,\,\pm\,1$$
\end{Theorem}
\noindent Olmsted's theorem is also proven independently by Jack S. Calcut in a more recent American Mathematical Monthly \cite{CAL} using Gaussian integers.

We also recall the following classical result.
\begin{Lemma}
If $\frac{\theta}{\pi}$ is irrational, then the sequence $(e^{in\theta})_{n\geq 0}$ is dense in the unit circle.
\end{Lemma}
We will use Olmsted's Theorem to show that our phase $\theta$ is irrational in degrees. We show a more general result.
\begin{Lemma}
Let $p$ be a prime number and $a$ and $b$ be any given two non-zero rational numbers such that
$$\left\lbrace\begin{array}{l}a^2\neq 1+pb^2\hfill\qquad (\star)\\a\neq \pm 1\pm\sqrt{2+pb^2}\hfill \qquad(\star\star)\end{array}\right.$$
Then, $ArcTan(a+b\sqrt{p})$ is irrational in degrees.
\end{Lemma}
\textsc{Proof.} First we show that at least one of $$ArcTan(a+b\sqrt{p})\;\text{ and }\;ArcTan(a-b\sqrt{p})$$ must be irrational in degrees. We then deduce that both numbers are in fact irrational in degrees.\\
Our argument is based on three ingredients: the contrapositive of Olmsted's theorem, an elementary equality giving the tangent of a sum and elementary facts about Galois conjugates. \\\\\textbf{Notation}. If $x=a+b\sqrt{p}$, denote by $\bar{x}$ the conjugate of $x$ that is,
$\bar{x}=a-b\sqrt{p}$. \\\\We will use that
$$\begin{array}{l}
\text{(i) $x+\bar{x}$ and $x\bar{x}$ are both rationals. }\\
\text{(ii) The sum (resp product) of two conjugates is the conjugate of the sum} \\\text{(resp product). }
\end{array}$$
A starting point is the classical congruence
\begin{equation}
ArcTan(x)+ArcTan(y)\equiv ArcTan\bigg(\frac{x+y}{1-xy}\bigg)\qquad\qquad\textit{Modulo $\pi$}
\end{equation}
Use congruence $(2.2)$ with $x=a+b\sqrt{p}$ and $y=\bar{x}$.
By the contrapositive of Olmsted's Theorem, if $r$ is a rational number distinct from $0,-1,1$, then $ArcTan(r)$ is irrational in degrees.
The assumption $(\star\star)$ on $a$, $b$ and $p$ guarantees that
$$\frac{2a}{1-a^2+pb^2}\not\in\lbrace 0,-1,1\rbrace$$
Then the sum
$$ArcTan(x)+ArcTan(\bar{x})$$ is irrational in degrees. And so at least one of $ArcTan(x)$, $ArcTan(\bar{x})$ must be irrational in degrees.
Without any loss of generality, suppose $ArcTan(x)$ is irrational in degrees. If $ArcTan(\bar{x})$ were rational in degrees,
then we could write
$$ArcTan(\bar{x})=\frac{p\pi}{q},$$
some $p\in\mathbb{Z}$, $q\in\mathbb{N}-\lbrace 0\rbrace$ with $p\wedge q=1$.
Then, \begin{equation}Tan(q\,ArcTan(\bar{x}))=0\end{equation}
Now use
\begin{equation}
Tan(x+y)=\frac{Tan(x)+Tan(y)}{1-Tan(x)Tan(y)}
\end{equation}
in order to prove by induction on the integer $q\geq 2$ the following claim.
\newtheorem{Claim}{Claim}
\begin{Claim}
$Tan(q\,ArcTan(x))\in\mathbb{Q}(\sqrt{p})$ and
$$\overline{Tan(q\,ArcTan(x))}=Tan(q\,ArcTan(\bar{x}))$$
\end{Claim}
\textsc{Proof.} We have
$$Tan(2ArcTan(x))=\frac{2x}{1-x^2}\in\mathbb{Q}(\sqrt{p})$$
and \begin{eqnarray}\overline{Tan(2ArcTan(x))}&=&\frac{2\bar{x}}{1-\bar{x}^2}\\
&&\notag\\
&=&Tan(2ArcTan(\bar{x}))\end{eqnarray}
where Eq. $(2.5)$ follows from point $(ii)$ above. So the claim holds for $q=2$.
Let $q\geq 3$ and suppose the claim holds for $q-1$. Write
$$Tan(q\,ArcTan(x))=\frac{Tan((q-1)ArcTan(x))+x}{1-x\,Tan((q-1)ArcTan(x))}$$
By induction hypothesis and point $(ii)$, the claim obviously holds for $q$ and so it holds for all $q\geq 2$. \\\\
Eq. (2.3) and Claim $1$ imply that
$$Tan(q\,ArcTan(x))=0$$
And so we have $$q\,ArcTan(x)\in\mathbb{Z}\pi$$
In other words, $ArcTan(x)$ is rational in degrees, a contradiction. \\


Below, we draw a summary of one possible complete protocol which realizes the ancilla $A_f$.
\begin{center}
\epsfig{file=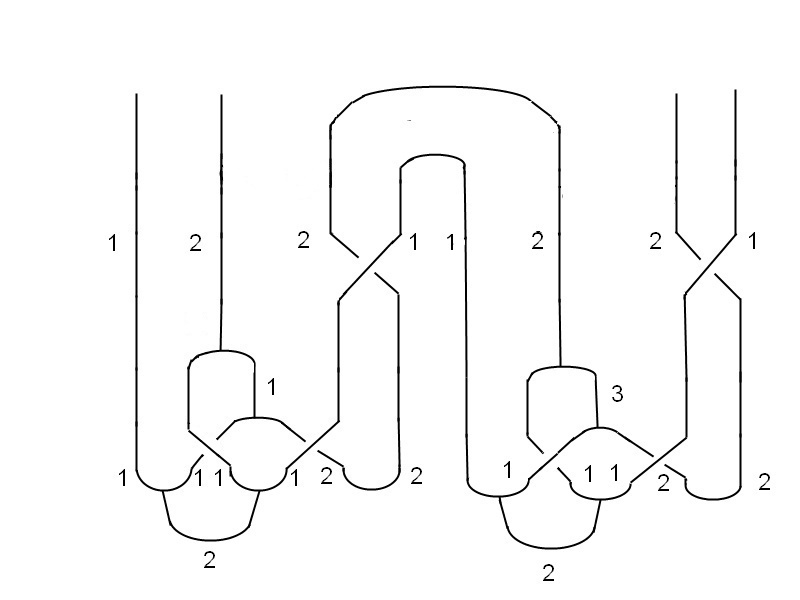, height=9cm}
\end{center}
\section{Irrational qutrit phase gate}
Again, the phase gate is built by ancilla preparation, then fusion of the ancilla into the input to form the gate. The second part appears in Chapter $1$. Only the ancilla preparation is detailed here.\\
Like announced and explained in the introduction of the paper, the ancilla preparation relies crucially on three different lemmas whose statements and proofs appear below.
\subsection{Qutrit projection}
The goal here is the following. Given any qutrit $a_0|0>+a_2|2>+a_4|4>$, find a protocol which projects this qutrit onto either the first two trits or the last two trits, up to some changes in the amplitudes.
\begin{Lemma}
The respective protocols below project the qutrit onto either $|0>$ and $|2>$ (protocol to the left) or $|4>$ and $|2>$ (protocol to the right).
\begin{center}
\epsfig{file=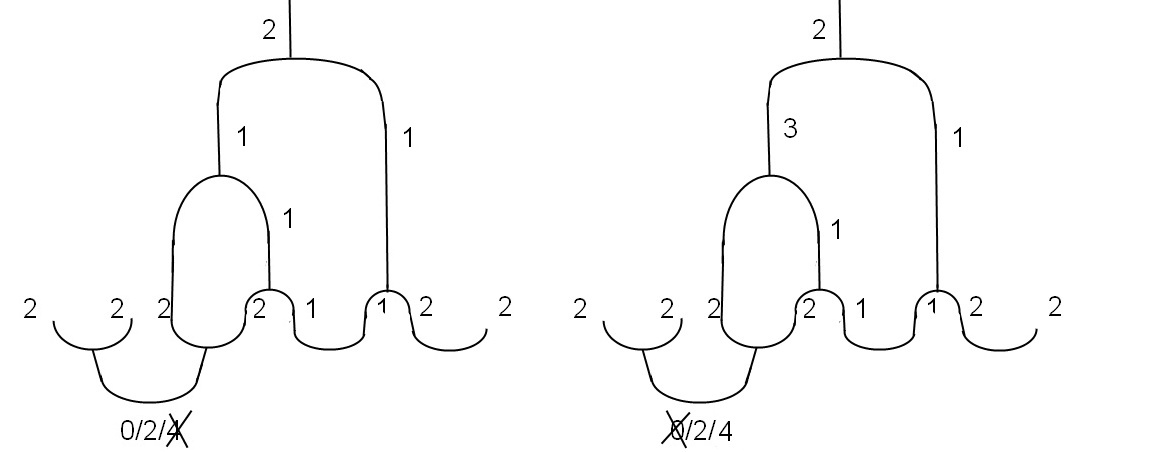, height=6cm}
$$\begin{array}{cc}a_0|0>+a_2|2>+a_4|4>\;\longrightarrow\;a_0|0>+\frac{1}{2}\,a_2|2>
&a_0|0>+a_2|2>+a_4|4>\;\longrightarrow\;a_4|4>+\frac{1}{2}\,a_2|2>\end{array}$$
\end{center}
\end{Lemma}

The idea is to get rid of one of the $|0>$ or $|4>$ contributions by fusion. Recall that
$$\left\lbrace\begin{array}{l}\text{$1$ and $0$ can fuse into $1$}\\
\text{$1$ and $4$ cannot fuse into $1$}\end{array}\right.$$
Also,
$$\left\lbrace\begin{array}{l}\text{$1$ and $4$ can fuse into $3$}\\
\text{$1$ and $0$ cannot fuse into $3$}\end{array}\right.$$
So bring a pair $1$'s out of the vacuum and do two fusions like on the figures.
\begin{center}
\epsfig{file=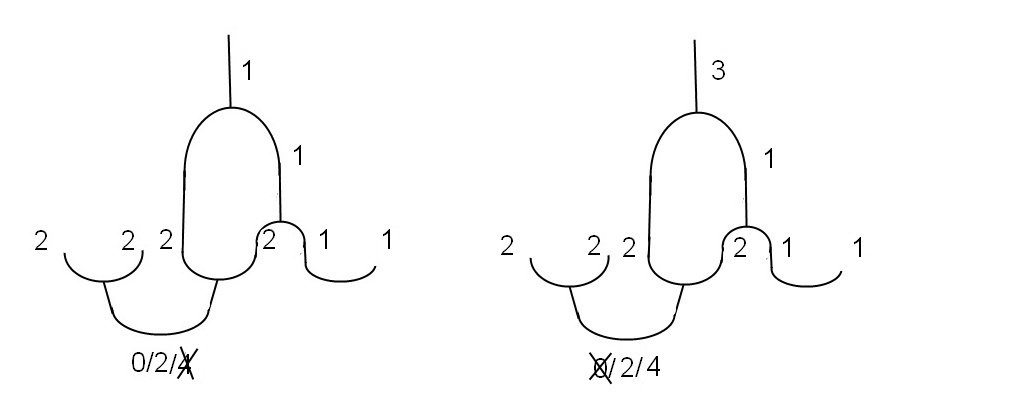, height=5cm}
\end{center}
\noindent The second fusion outcome determines which contribution $|0>$ or $|4>$ vanishes. By the "triangle collapse" move (cf Fig. $1$), the fusion rules must indeed be respected between $1$, the qutrit and $1$ (or $3$) depending on the outcome.
In order to get back to a qutrit $2222$, it will suffice to bring a pair of $2$'s and do additional fusions.
\begin{center}
\epsfig{file=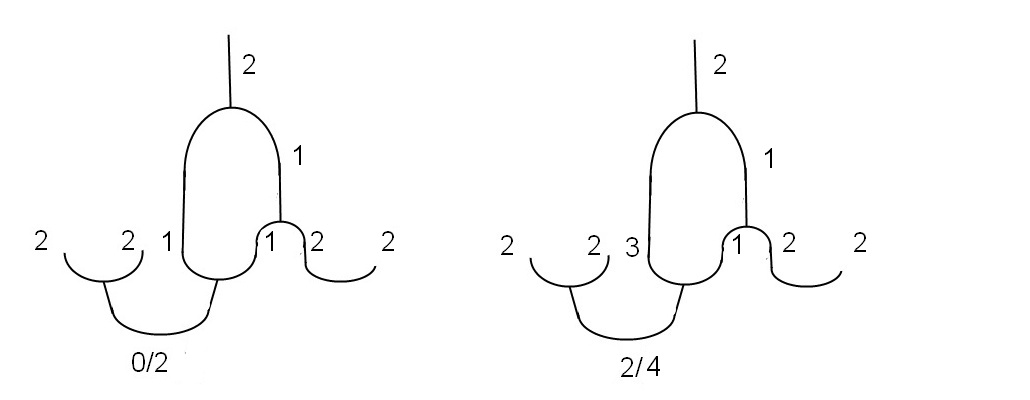, height=5cm}
\end{center}
Regarding the coefficients, the only relevant symbols are
\begin{itemize}
\item For the first part, $$\left\lbrace\begin{array}{ccc}2&2&1\\1&1&2\end{array}\right\rbrace^u=\frac{1}{\sqrt{2}},\;
\left\lbrace\begin{array}{ccc}2&2&3\\1&1&2\end{array}\right\rbrace^u=-\frac{1}{\sqrt{2}},\;\left\lbrace\begin{array}{ccc}4&2&3\\1&1&2\end{array}
\right\rbrace^u=1$$
\item For the second part, $$\left\lbrace\begin{array}{ccc}2&1&2\\1&2&1\end{array}\right\rbrace^u=\frac{1}{\sqrt{2}},\; \left\lbrace\begin{array}{ccc}2&3&2\\1&2&1\end{array}\right\rbrace^u=-\frac{1}{\sqrt{2}},\;\left\lbrace\begin{array}{ccc}4&3&2\\1&2&1\end{array}\right\rbrace^u=1$$
\end{itemize}
The next step allows to fuse together two qutrits that have previously been projected to $|0>$ and $|2>$ (or to $|4>$ and $|2>$).
\subsection{Qutrit fusion}
The idea underlying this fusion is to keep the $(|0>,|2>)$ contribution and the $(|2>,|0>)$ contribution, having in mind that if one ancilla is the conjugate of the other one, then we will get equal norms in $|0>$ and $|2>$ and twice the interesting relative phase we had. Thus, we want to fuse both ancillas into a $2$, that is fuse anyons $4$ and $5$ to $0$, then fuse anyons $3$ and $4$ to $2$. The issue of the $(|2>,|2>)$ contribution gets resolved by the fact that, specific to this theory, the $6j$-symbols with only $2$'s in it is zero, hence this contribution vanishes during the process.
The second magic thing about this fusion is that when we fuse again in order to retrieve a qutrit $2222$, we create the desired factor of $\frac{1}{\sqrt{2}}$.
\begin{Lemma} The following protocol realizes the fusion of two "projected" ancillas into a third ancilla, where the amplitudes have been multiplied term by term like indicated below, and where a factor $\frac{1}{\sqrt{2}}$ has been generated in front of the $|2>$ trit.
\begin{center}
\epsfig{file=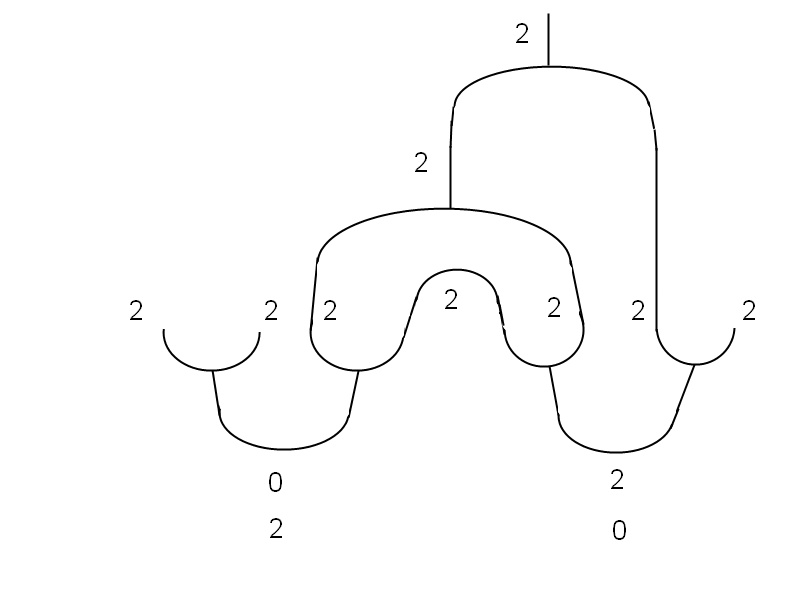, height=6cm}
\end{center}
$$\begin{array}{cccccccc}
\begin{array}{l}|0>\\|2>\\|4>\end{array}&\nts\nts\nts\begin{pmatrix}a_0\\a_2\\0\end{pmatrix}&,&
\begin{array}{l}|0>\\|2>\\|4>\end{array}&\nts\nts\nts\begin{pmatrix}b_0\\b_2\\0\end{pmatrix}&\overset{Fusion}{\longrightarrow}&
\begin{array}{l}|0>\\|2>\\|4>\end{array}&\nts\nts\nts\begin{pmatrix}a_0b_2\\\frac{1}{\sqrt{2}}a_2b_0\\0\end{pmatrix}
\end{array}$$
\end{Lemma}
\textsc{Proof.} The proof is straigtforward. The factor $\frac{1}{\sqrt{2}}$ arises from the $6j$-symbol
$$\left\lbrace\begin{array}{ccc}2&2&2\\2&2&0\end{array}\right\rbrace^u=\frac{1}{\sqrt{2}}$$\hfill$\square$\\
Here is how we apply the lemma.
Suppose we know how to prepare two ancillas
$$\begin{array}{ccc}\begin{array}{cc}\begin{array}{l}|0>\\|2>\\|4>\end{array}&\nts\nts\nts\begin{pmatrix}a\\\frac{1}{2}b\\0\end{pmatrix}\end{array}&\text{and}&
\begin{array}{cc}\begin{array}{l}|0>\\|2>\\|4>\end{array}&\nts\nts\nts\begin{pmatrix}\bar{a}\\\frac{1}{2}\bar{b}\\0\end{pmatrix}\end{array}\end{array}$$

\noindent with relative phase $\theta$ between complex numbers $a$ and $b$. Then, by fusion, we are able to make a third ancilla
$$\begin{array}{cc}\begin{array}{l}|0>\\|2>\\|4>\end{array}&\nts\nts\nts\begin{pmatrix}a\bar{b}\\\frac{1}{\sqrt{2}}b\bar{a}\\0\end{pmatrix}\end{array}$$
which has the norms we want and an interesting relative phase of $2\theta$.
\subsection{Qutrit pair fusion}
The next protocol allows, given a qutrit ancilla, to get another ancilla which is obtained from the first one by shifting and cloning its $|2>$ contribution to the $|0>$ and $|4>$ trits and by adding its $|0>$ and $|4>$ contributions to yield the $|2>$ contribution of the new ancilla.
\begin{Lemma} Pair of $2$'s fusion into the qutrit $2222$.
\begin{center}
\hspace{-0.4cm}\epsfig{file=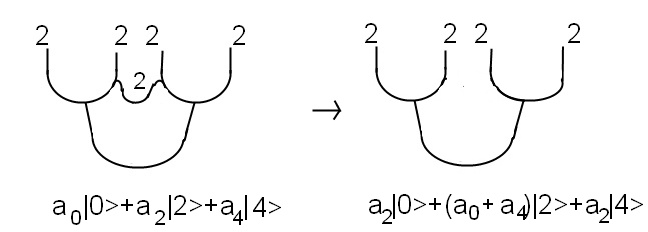, height=5cm}
\end{center}
\end{Lemma}
\textsc{Proof.} The proof can be read out of the following figure.\\
\begin{center}
\epsfig{file=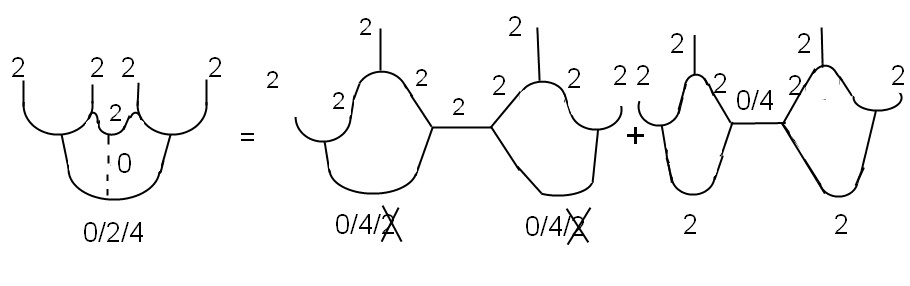, height=3.7cm}
\end{center}
We do an F-move on the diagram to the left on the zero labeled edge. Then, on the first diagram of the sum, the $|2>$ trit contribution vanishes. Indeed, if we attempt to remove the bubble, the $6j$-symbol arising during the removal process is
$$\left\lbrace\begin{array}{ccc}2&2&2\\2&2&2\end{array}\right\rbrace$$
which is zero in this theory. Up to an overall scalar, we get
$$\nts\nts\nts\nts\nts\nts\nts\nts\nts\nts\nts\begin{array}{cc}\begin{array}{l}|0>\\\\\\|2>\\\\\\|4>\end{array}&\nts\nts\nts\begin{pmatrix}a_2\bigg(\left\lbrace\begin{array}{ccc}2&2&2\\2&0&2\end{array}
\right\rbrace^u\bigg)^2\left\lbrace\begin{array}{ccc}2&2&0\\2&2&0\end{array}
\right\rbrace^u\\\\a_0\bigg(\left\lbrace\begin{array}{ccc}2&2&2\\2&2&0\end{array}
\right\rbrace^u\bigg)^2\left\lbrace\begin{array}{ccc}2&2&2\\0&0&0\end{array}
\right\rbrace^u+a_4\bigg(\left\lbrace\begin{array}{ccc}2&2&2\\2&2&4\end{array}
\right\rbrace^u\bigg)^2\left\lbrace\begin{array}{ccc}2&2&2\\4&4&0\end{array}
\right\rbrace^u\\\\a_2\bigg(\left\lbrace\begin{array}{ccc}2&2&2\\2&4&2\end{array}
\right\rbrace^u\bigg)^2\left\lbrace\begin{array}{ccc}2&2&4\\2&2&0\end{array}
\right\rbrace^u\end{pmatrix}\end{array}$$

The squared values or values of the $6j$-symbols are in the same order as in which they appear,
\begin{center}
$1,\frac{1}{2}$\\
$\;$\\
$\frac{1}{2},1,\frac{1}{2},1$\\
$\;$\\
$1,\frac{1}{2}$
\end{center}

Hence we get an overall $\frac{1}{2}$ and the coefficients announced. \hfill $\square$\\\\
If we apply this protocol of pair fusion to the ancilla of last section, here is what we get
$$\begin{array}{cc}\begin{array}{l}|0>\\|2>\\|4>\end{array}&\nts\nts\nts\begin{pmatrix}a\bar{b}\\\frac{1}{\sqrt{2}}b\bar{a}\\0\end{pmatrix}\end{array}
\overset{\text{After pair fusion}}{\longrightarrow} \begin{array}{cc}\begin{array}{l}|0>\\|2>\\|4>\end{array}&\nts\nts\nts\begin{pmatrix}\frac{1}{\sqrt{2}}b\bar{a}\\a\bar{b}\\\frac{1}{\sqrt{2}}b\bar{a}
\end{pmatrix}\end{array}$$
It remains to construct an ancilla
$$\begin{array}{cc}\begin{array}{l}|0>\\|2>\\|4>\end{array}&\nts\nts\nts\begin{pmatrix}a\\b\\0\end{pmatrix}\end{array}$$ and its conjugate ancilla with interesting relative phase between the two complex numbers $a$ and $b$. This is the object of the next part.
\subsection{Final ancilla preparation}
We start with the same configuration as for the qubit, that is four anyons $(1111)_2$ and a pair of $2$'s. We do the following braids and fusions.
\begin{center}
\epsfig{file=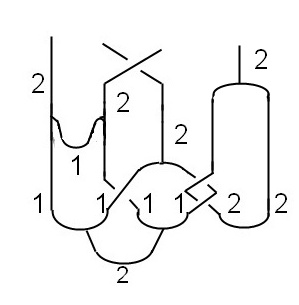, height=7cm}
\end{center}
The full twist on $1122$ swaps the $|0>$ and $|2>$ bits, see \cite{QIP}, hence creates a $2$ charge line between the qubit $1111$ and the pair of $2$'s. The single braid on $(1111)_2$ creates a superposition of $|0>$ and $|2>$. After the three fusions like on the picture, we get a qutrit $2222$, which is up to overall complex scalar
$$\sqrt{2}e^{-i\frac{\pi}{3}}|0>+\frac{1}{2}|2>$$
Of course the relative phase is not yet satisfactory, but our preparatory efforts will be rewarded by a $\sigma_2$-braid on that vector.
It yields, still up to overall complex scalar,
$$\begin{array}{cc}\begin{array}{l}|0>\\\\|2>\\\\|4>\end{array}&\nts\nts\nts\begin{pmatrix}\sqrt{7}e^{i\gamma^{'}}\\\\e^{i\frac{\pi}{4}}\\\\\sqrt{3}e^{i\frac{\pi}{12}}
\end{pmatrix}\end{array}\text{with $\gamma^{'}=\pi+ArcTan\bigg(\frac{-14+3\sqrt{3}}{13}\bigg)$}$$
Our Lemma $3$ does not apply here because condition $(\star)$ is not satisfied. \\However, we can still conclude by \cite{CAL2}. In \cite{CAL2}, J.S. Calcut determines all of the numbers $Tan(\frac{p\pi}{q})$ with degree two or less over $\mathbb{Q}$. He shows that for such numbers, the only possible quadratic irrational values are
 $$\left\lbrace\begin{array}{l}\pm\sqrt{3}\\\\\pm\frac{\sqrt{3}}{3}\\\\\pm 1\pm\sqrt{2}\\\\\pm 2\pm\sqrt{3}\end{array}\right.$$

\noindent Thus, the relative phase of our newly produced qutrit ancilla is irrational in degrees. Moreover, we claim that the same protocol where the two single braids of the picture have been replaced with their inverse braids as on the new picture below produces the conjugate ancilla. This is simply because each matrix corresponding to a $\sigma_2$ action on $1111$ and $2222$ is self-transpose, so that the inverse matrices are the conjugate matrices. We obtain the conjugate ancilla
$$\begin{array}{cc}\begin{array}{l}|0>\\\\|2>\\\\|4>\end{array}&\nts\nts\nts\begin{pmatrix}\sqrt{7}e^{-i\gamma^{'}}\\\\e^{-i\frac{\pi}{4}}\\\\\sqrt{3}e^{-i\frac{\pi}{12}}
\end{pmatrix}\end{array}$$
\begin{center}
\epsfig{file=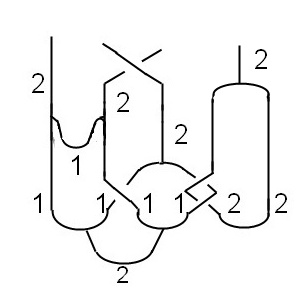, height=7cm}
\end{center}
By applying a qutrit projection on both ancillas, then a qutrit fusion of the two ancillas, followed by one qutrit pair fusion, we are able to make the final ancilla, say $B_f$, which is up to overall complex scalar,
$$\boxed{B_f=\begin{array}{ccc}\begin{array}{cc}\begin{array}{l}|0>\\|2>\\|4>\end{array}&\nts\nts\nts\begin{pmatrix}1\\\sqrt{2}\,e^{i\gamma}\\1\end{pmatrix}\end{array}
&\text{with}&\gamma=2\,ArcTan\bigg(\frac{3\sqrt{3}-14}{13}\bigg)-\frac{\pi}{2}\end{array}}$$
By Chapter $1$, we are able to turn this ancilla into an irrational phase gate
$$\begin{pmatrix} 1&&\\&e^{i\gamma}&\\&&1\end{pmatrix}$$
Below, we drew a picture which assembles all the different parts of the protocol together to produce the ancilla $B_f$.
\begin{center}
\epsfig{file=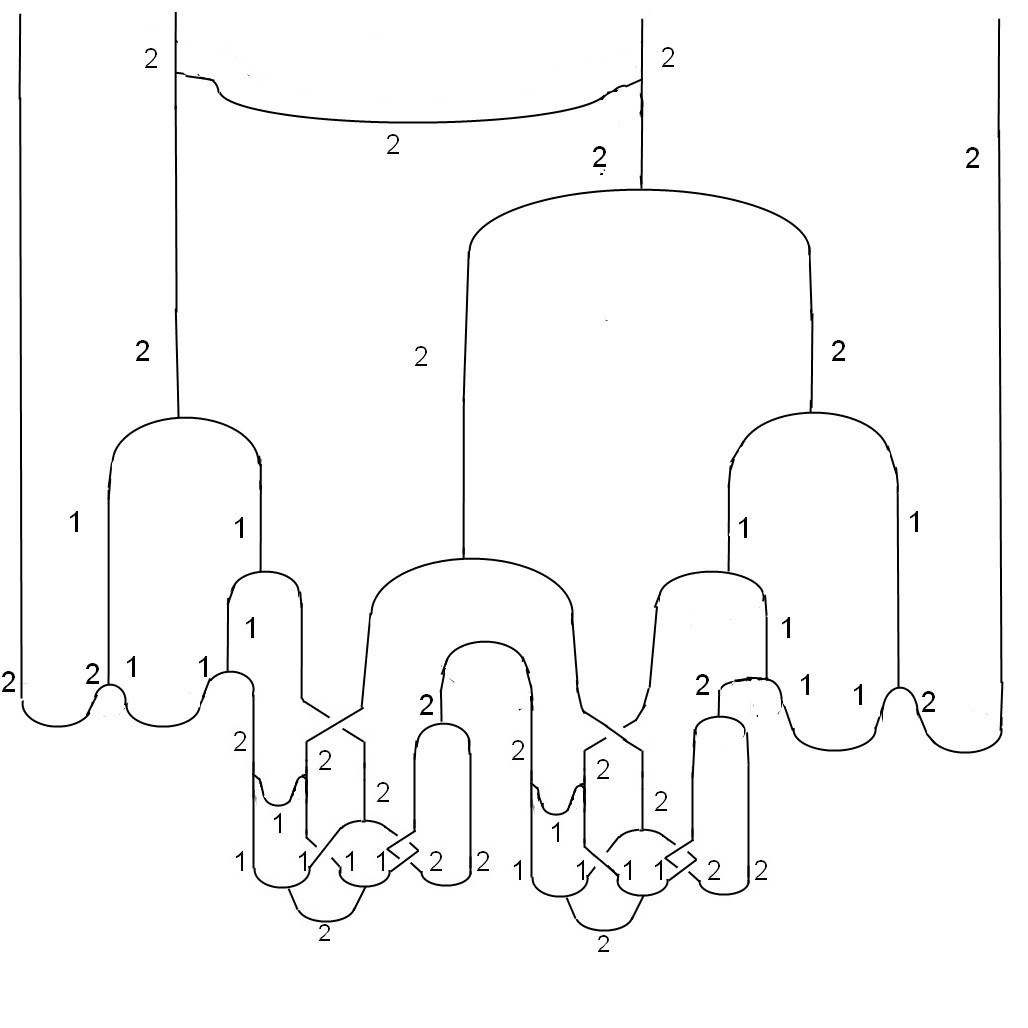, height=12cm}
\end{center}
To conclude, we comment on the similarities and differences between this ancilla preparation and the one we did for the quantum bit. We only focus on the first step of each protocol, namely the step that yields the "interesting relative phase" and whose starting point is in both cases the qubit $1111$ and a pair of $2$'s.

For the qubit, the information is carried ultimately by the edge between the qubit $1111$ and the ancilla pair $22$, whereas for the qutrit, the information is carried by the edge from the qubit $1111$.

Another aspect is that for the qubit the interesting phase is really arising from one bit and one bit only (either $|1>$ or $|3>$) as a sum of two contributions carried by the edge of the qubit $1111$. Similarly for the qutrit, the interesting phase is carried by one trit only, namely the $|0>$ trit, after the braiding of the superposition of $|0>$ and $|2>$ has occurred. There would have been a priori two candidates, namely $|0>$ or $|4>$ (it would not be $|2>$ by the shape of the $\sigma_2$ matrix possessing a zero in its central position), but by the game of the coefficients and phases, it is the $|0>$ trit which will carry the irrational phase.

Finally, in the protocol for the qutrit, the pair fusion is crucial, whereas in the protocol for the qubit,
we only do simple fusions (this is allowed by adjusting the outcome from the first fusion, thus avoiding the Freedman pair of $4$'s fusion).

\begin{center}
\textbf{Acknowledgements}
\end{center}
We are pleased to thank Bela Bauer, Stephen Bigelow, Parsa Bonderson, Meng Cheng, Spiros Murakalis, Chetan Nayak and Zhenghan Wang for helpful discussions.

\chapter{Realizing the $2$-qubit permutation gates}
\begin{center}
\textbf{Abstract}\end{center}
We give a protocol which generates one or the other of two $2$-qubit entangling gates. By acting on these two gates with one of the $2$-qubit entangling gates that was recently made in \cite{QIP} and by additional braiding, we are able to make the gate
$$CNOT.\,SWAP=\begin{pmatrix}1&0&0&0\\0&0&1&0\\0&0&0&1\\0&1&0&0\end{pmatrix}$$ Since the SWAP gate is easily realized by braiding only, we are thus able to make the CNOT gate. Further, with additional pair fusion operations, we are able to generate all the permutation gates of $Sym(4)$.

\section{Introduction}
The Brylinski couple, in a work \cite{BRY} dating from $2001$, proves that universal single qudit gates together with any $2$-qudit entangling gate are sufficient for universal quantum computation. For qubits, there is an independent proof of this fact by \cite{BCD}. In \cite{CL2}, we give a protocol which produces a $2$-qubit entangling gate whose definition appears in Theorem $1$ of the paper. With additional braiding, we also make a circulant $2$-qubit entangling gate. The goal of the present paper is to make all the $2$-qubit quantum gates from $Sym(4)$ in the Kauffman-Jones version of $SU(2)$ Chern-Simons theory at level $4$. Our qubit is formed by four anyons of topological charges $1221$, just like in \cite{CL2}. The difficult part is to make a $3$-cycle. For instance the products of two transpositions with disjoint support are easily obtained by some Freedman type \cite{CL} fusion operations. The main protocol developed in our paper relies on two novel ideas, which we name "qubit transfer" and "qubit demolition".
First we move the information carried by the right qubit, second we destroy the right qubit without leaking any quantum information. During the second phase, namely the demolition phase, we use a process that was described in \cite{CL2} as "Process (P)" which operates a projection for a qutrit $0/2/4$ onto either $2$ or $0/4$. For each of both outcomes of the projection, we get a distinct gate. By acting on each of these two gates to the left by
$$\begin{array}{cc}&\begin{array}{cccc}|11>&|13>&|31>&|33>\end{array}\\\begin{array}{l}|11>\\\\|13>\\\\|31>\\\\|33>\end{array}&\begin{pmatrix}
-\frac{1}{2}&0&0&-i\frac{\sqrt{3}}{2}\\\\0&-\frac{1}{2}&-i\frac{\sqrt{3}}{2}&0\\\\
0&-i\frac{\sqrt{3}}{2}&-\frac{1}{2}&0\\\\-i\frac{\sqrt{3}}{2}&0&0&-\frac{1}{2}\end{pmatrix}\end{array}=AUX$$

\noindent a $2$-qubit entangling gate that was produced in \cite{CL2}, and together with some additional braiding, we are able to always make the CNOT.SWAP gate independently from the resulting outcome of Process (P).

While by the Brylinski result, only one $2$-qubit entangling gate like built in \cite{CL2}, together with a universal set of $1$-qubit gates like resulting from Chapter $2$ would be sufficient in order to approximate any permutation gate of $Sym(4)$, one of the highlights of our paper is to give a way to physically make these permutation gates accurately instead of simply by approximation.

Throughout the paper, we assume that the reader is familiar with the Kauffman-Jones theory and a nice reference for it is \cite{KL}. For the unitary version of this theory, a good reference is \cite{ZW}\cite{CL}. We also assume that the reader is familiar with interferometric measurements and good references are \cite{BO}\cite{BO2}\cite{BO3}\cite{FL}.

Our main protocol uses braids, interferometric measurements, fusions and unfusions of anyons, vacuum pair creation and recovery procedures, ideas which originated in \cite{MO}.

\section{Main protocol}
The protocol is based on qubit transfer, that is the qubit outputs can be different from the qubit inputs. Which means a second part of the protocol must deal with qubit demolition: after the information has been transferred onto another qubit, the old qubit must be destroyed without leaking any quantum information. In the protocol which we present, the left qubit input is also the left qubit output, but the right qubit input is not the right qubit output and must be destroyed. The ancillas are simply pairs created out of the vacuum which help with the destruction process as well as with the final separation process. Whereas, in the protocol of \cite{CL2}, there are two qubit ancillas which become the qubit outputs.
We provide the protocol, then explain it. As part of the protocol, some recovery procedures are needed and get provided in the discussion which follows the theorem. In order to give a figure that is readable, the recovery procedures don't appear in the figure, but are nevertheless a crucial part of the protocol. Thus, the figure only provides a winning protocol, that is one with favorable measurement outcomes. The current section deals with the statement of  Theorem $3$ and its proof.
\begin{Theorem}
The following protocol
\begin{center}
\epsfig{file=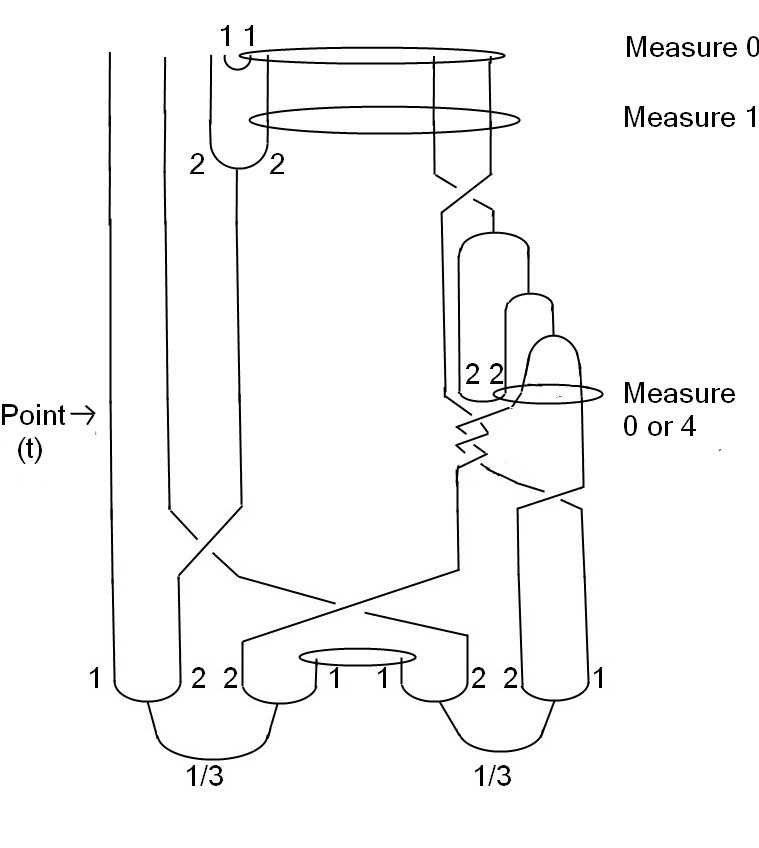, height=13cm}
\end{center}
produces the entangling gate
$$EG^{(1)}=\begin{array}{cc}\begin{array}{l}\\\\|11>\\\\|13>\\\\|31>\\\\|33>\end{array}&\nts\nts\begin{array}{l}\begin{array}{cccc}\;\;\;|11>&\;|13>&\;\;\;|31>&\;\;\;\nts|33>\end{array}\\\\
\begin{pmatrix}1/4&i\sqrt{3}/4&3/4&-i\sqrt{3}/4\\&&&\\3/4&-i\sqrt{3}/4&1/4&i\sqrt{3}/4\\&&&\\
i\sqrt{3}/4&-3/4&-i\sqrt{3}/4&-1/4\\&&&\\-i\sqrt{3}/4&-1/4&i\sqrt{3}/4&-3/4\end{pmatrix}\end{array}\end{array}$$
The next protocol
\begin{center}
\epsfig{file=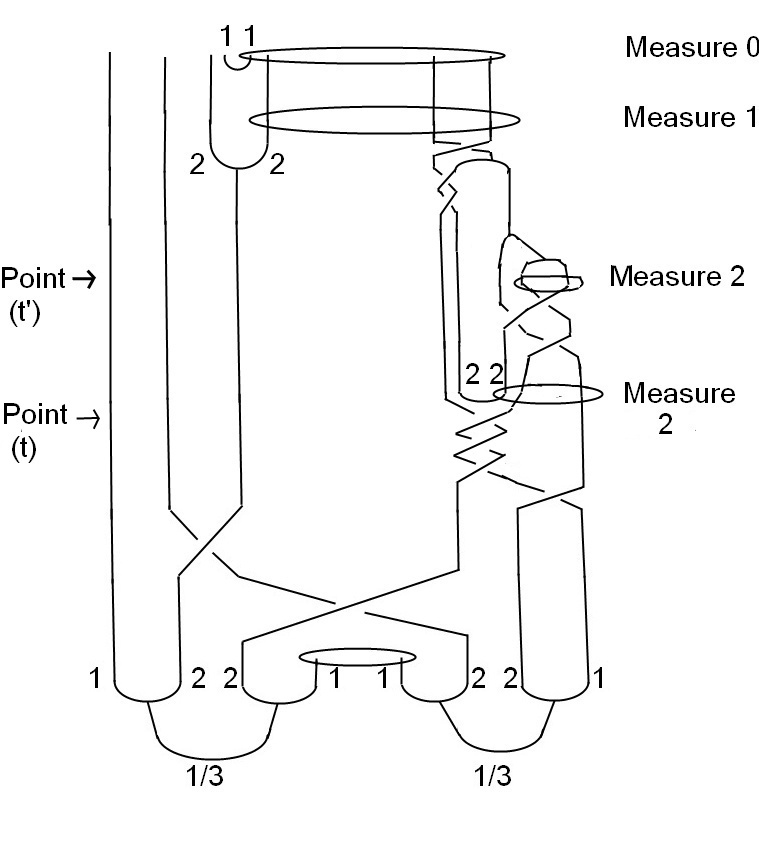, height=13cm}
\end{center}
produces the entangling gate
$$EG^{(2)}=\begin{array}{cc}\begin{array}{l}\\\\|11>\\\\|13>\\\\|31>\\\\|33>\end{array}
&\nts\nts\begin{array}{l}\begin{array}{cccc}\;\;\;|11>&\;|13>&\;\;\;|31>&\;\;\;\nts|33>\end{array}\\\\
\begin{pmatrix}1/4&-i\sqrt{3}/4&3/4&i\sqrt{3}/4\\&&&\\3/4&i\sqrt{3}/4&1/4&-i\sqrt{3}/4\\&&&\\
i\sqrt{3}/4&3/4&-i\sqrt{3}/4&1/4\\&&&\\-i\sqrt{3}/4&1/4&i\sqrt{3}/4&3/4\end{pmatrix}\end{array}\end{array}$$

\noindent\textbf{Recovery:} if at point $(t^{'})$, the measurement rather produces the outcome $0$ (resp $4$) instead of $2$, then remove the pair of $2$'s composed by the last two anyons (resp fuse the last three anyons two by two from the right, then fuse a pair of $4$'s in the last two anyons). By doing so, get back to a previous stage in time, namely to point $(t)$. Proceed again from there and generate either the first gate of the Theorem or the second gate depending on the subsequent measurement outcomes.
\end{Theorem}

\textsc{Proof of Theorem $3$.} \textbf{Unless otherwise mentioned in the discussion, it is understood that we deal with the first protocol.} \\

The protocol can be divided into a few main steps. \\

\textit{Step $1$}. Measure anyons $4$ and $5$ to zero in order to fuse them (we refer the reader to Chapter $1$ for the recovery to follow when the outcome of this first measurement is not zero).\\

\textit{Step $2$}. The first two braids of the picture are the entangling braids. The $2$-qubit output is carried by the edge from the left qubit input and the edge from the fusion of Step $1$. It will remain to destroy the right qubit input independently from what the $2$-qubit input was, then get back to a $2$-qubit shape.\\

\textit{Step $3$}. The qubit demolition is made possible by a series of two braids, followed by a measurement assisted by a pair of $2$'s out of the vacuum.\\

\textit{Step $4$}. All the steps which follow Step $3$ deal with the "back into shape" process, that is get back to a $2$-qubit. This step is made possible by three fusions, one braid, one unfusion, an interferometric measurement, adding a pair out of the vacuum and running a second interferometric measurement. Though this seems like a long and complicated process, all these steps arise "naturally" from reading the figure. Thus, the key ideas of this protocol remain the qubit transfer and the qubit demolition.\\

We will go over each step, except Step $1$ which is extensively detailed in Chapter $1$. In order to understand the entangling braids, we will need to know the $\sigma_2$-actions on the qubits $1221$, $3223$, $1223$ and $3221$. The first two actions and the last two are identical, hence there are only two matrices of interest here. We have
$$\begin{array}{l}G_2(1,2,2,1)=G_2(3,2,2,3)=\begin{pmatrix}-\frac{1}{2}&i\frac{\sqrt{3}}{2}\\&\\i\frac{\sqrt{3}}{2}&-\frac{1}{2}
\end{pmatrix}\\\\G_2(1,2,2,3)=G_2(3,2,2,1)=\begin{pmatrix}-i\frac{\sqrt{3}}{2}&\frac{1}{2}\\&\\\frac{1}{2}&-i\frac{\sqrt{3}}{2}\end{pmatrix}
\end{array}$$
And so by reading the first columns of the matrices above, the matrix representing the action of the first braid
\begin{center}\epsfig{file=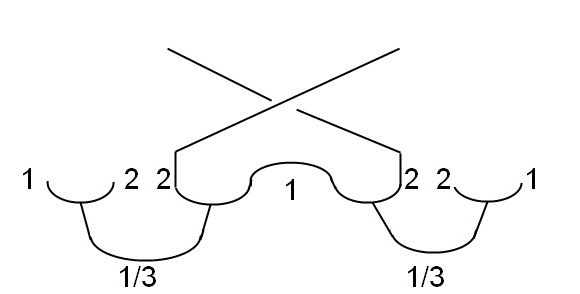, height=4cm}
\end{center}
is the following
\begin{center}
\epsfig{file=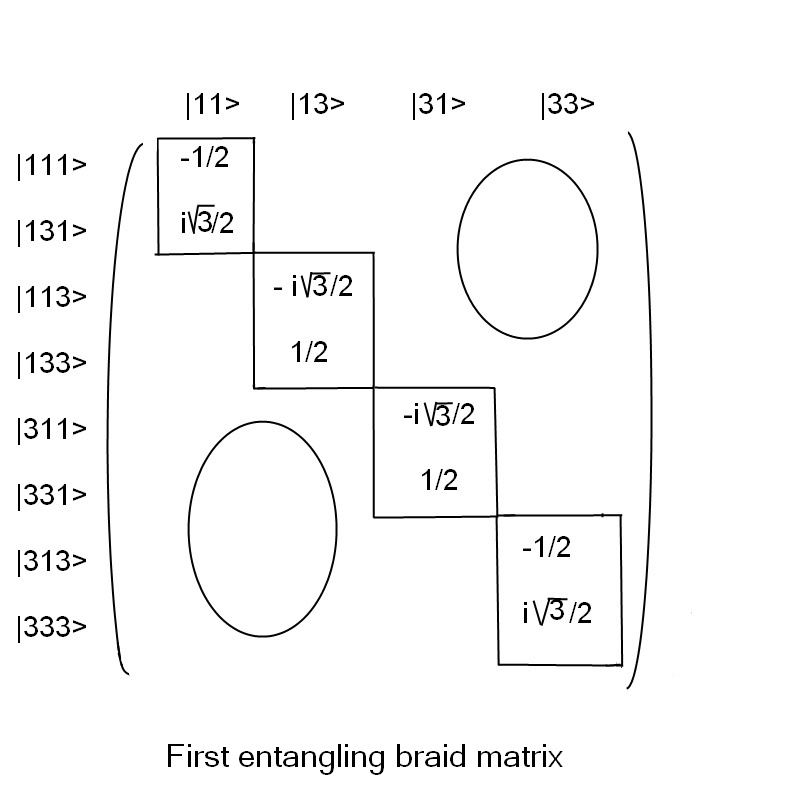, height=11cm}
\end{center}
Each row of the matrix is indexed in the same order as the horizontal edges appearing from left to right on the figure of the protocol. So that the first and last bits correspond to the input bits and the middle bit corresponds to the label carried by the middle edge after the braid has occurred. If the first and last bits are identical, then we read the superposition of the middle edge on the first column of the first two by two matrix above. When the first and last bits are distinct, the superposition is given by the first column of the second two by two matrix above.

Now the second braid. We braid $1221$ or $1223$ on the left input depending on whether the middle horizontal edge carries the label $1$ or the label $3$. We get the following superposition. For clarity we ordered the $2$-qubit input states differently, namely we permuted $|13>$ and $|31>$. We kept the same ordering as before on the rows. Since the right qubit input never gets braided, its label remains what it is as an input.
Thus, if the right qubit input is $1$, the last four rows of the matrix are filled with zeroes and if the right qubit input is $3$, the first four rows of the matrix are filled with zeroes. This motivated our choice of reordering the input basis. The reader gets invited to compute some of the coefficients provided in the matrix below in order to grasp the entangling action.
\begin{center}
\epsfig{file=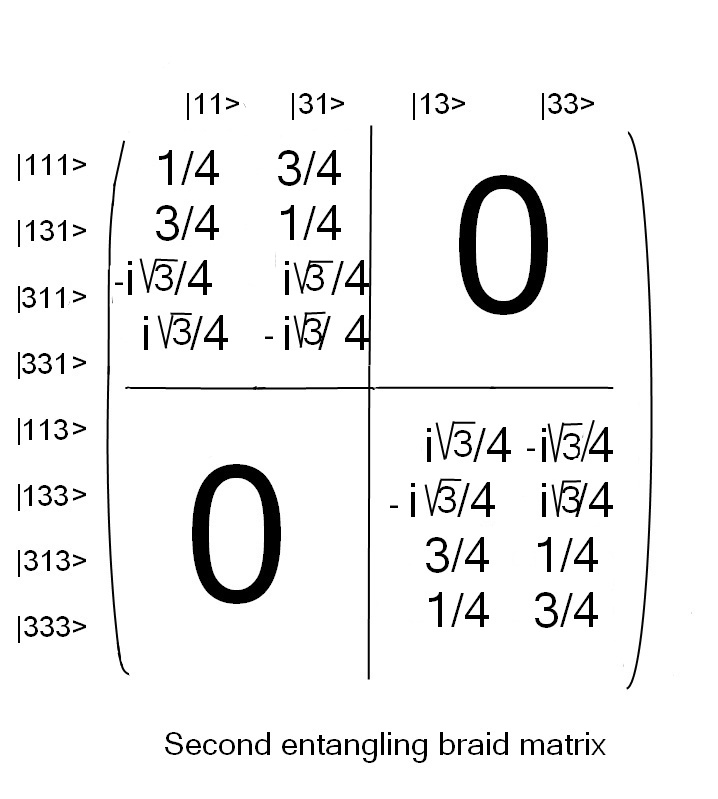, height=11cm}
\end{center}
The goal now is to get to a four by four matrix with the $2$-qubit output read out of the first two bits, that is find a way to suppress the right qubit. 
The situation is complex since we have a superposition of two qubits, namely $1221$ and $3221$, but we developed a technique that deals with it. We will need the respective matrices from the following actions
\begin{center}
\epsfig{file=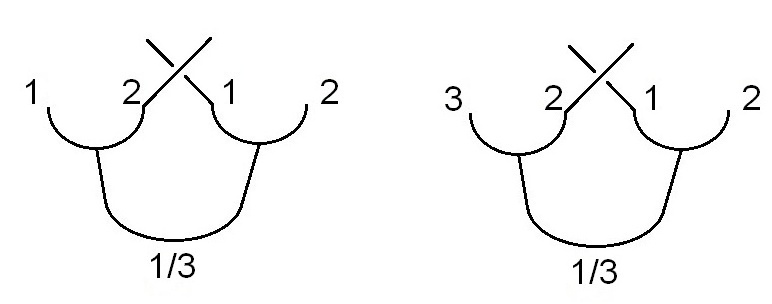, height=5cm}
\end{center}
These are
$$\begin{array}{cc}\begin{array}{cc}&\begin{array}{cc}|1>&|3>\end{array}\\&\\\begin{array}{l}|0>\\\\|2>\end{array}&\begin{pmatrix}
\frac{e^{i\frac{2\pi}{3}}}{\sqrt{2}}&\frac{e^{-i\frac{5\pi}{6}}}{\sqrt{2}}\\&\\
\frac{1}{\sqrt{2}}&-\frac{i}{\sqrt{2}}\end{pmatrix}\end{array}
&\begin{array}{cc}&\begin{array}{cc}|1>&|3>\end{array}\\&\\\begin{array}{l}|2>\\\\|4>\end{array}&\begin{pmatrix}
\frac{i}{\sqrt{2}}&\frac{1}{\sqrt{2}}\\&\\
\frac{e^{-i\frac{5\pi}{6}}}{\sqrt{2}}&\frac{e^{-i\frac{\pi}{3}}}{\sqrt{2}}
\end{pmatrix}\end{array}
\end{array}
$$
$$\begin{array}{l}\end{array}$$
Suppose we have put the right qubit under a superposition of shapes $1212$ and $3212$ by a single $\sigma_1$-braid exchanging the right most anyons of topological charges $1$ and $2$. The matrix of this exchange is
$$R(1,2)=\begin{pmatrix}
-e^{i\frac{\pi}{3}}&\\
&-e^{-i\frac{\pi}{6}}
\end{pmatrix}$$
The matrices gathering the two actions above read
$$\begin{array}{ccc}\begin{array}{ccc}&\begin{array}{cc}|1>&|3>\end{array}&\\&&\\\frac{1}{\sqrt{2}}&\begin{pmatrix}
1&1\\
-e^{i\frac{\pi}{3}}&e^{i\frac{\pi}{3}}\end{pmatrix}&\begin{array}{l}|0>\\|2>\end{array}\end{array}
&\begin{array}{ccc}&\begin{array}{cc}|1>&|3>\end{array}\\&&\\\frac{1}{\sqrt{2}}&\begin{pmatrix}
e^{-i\frac{\pi}{6}}&-e^{-i\frac{\pi}{6}}\\
i&i
\end{pmatrix}&\begin{array}{l}|2>\\|4>\end{array}\end{array}&(\text{\textbf{M}})
\end{array}
$$
$$\begin{array}{l}\end{array}$$

We then proceed after making a few observations. Those observations that got omitted are those which appear even more clearly on the matrices (M).\\
\begin{itemize}
\item For the $|2>$ outcome of the braid on $1212$ (second row of the left hand side matrix), the phases arising from both right input bits $|1>$ and $|3>$ are opposite and are, up to overall phase $-e^{i\frac{\pi}{3}}$, respectively $1$ and $-1$. \\
\item For the $|2>$ outcome of the braid on $3212$ (first row of the right hand side matrix), the phases arising from both right input bits $|1>$ and $|3>$ are opposite and are, up to overall phase $e^{-i\frac{\pi}{6}}$, respectively $1$ and $-1$. \\
\item Doing a full $\sigma_2$-twist on $1122$ swaps the bits $|0>$ and $|2>$ with overall phase $e^{i\frac{2\pi}{3}}$, see \cite{QIP}.\\
\item Likewise, doing a full $\sigma_2$-twist on $3122$ swaps the bits $|2>$ and $|4>$ with overall phase $-e^{i\frac{2\pi}{3}}$.
\end{itemize}
Suppose we do a series of operations consisting of doing the $\sigma_1$-braid, then doing the $\sigma_2$-braid on the superposition $1212$ and $3212$, then doing the full twist exchanging the bits $|0>$ and $|2>$ and $|2>$ and $|4>$ of the superposition,
followed by bringing a pair of $2$'s placed like on the main figure and measuring either $0$ or $4$ during the interferometric measurement of the main figure.
These outcomes have the effect of projecting the right qubit onto $2$ since $2$ and ($0$ or $4$) can only fuse into $2$.
Next, we fuse the anyons two by two from the right hand side until getting to a single anyon of topological charge $2$. Its left neighbor carries the topological charge $1$. Now interchange the relative positions of these two anyons by doing a $\sigma_1$-braid on them. By doing so, introduce a relative phase of $-i$ between the $|1>$ and the $|3>$ bits carried by the central horizontal edge. This relative phase of $-i$ compensates exactly the $i$ relative phase observed before between the two matrices $(M)$ after the projection onto $|2>$.  \\\\
\textbf{Thus, the only relative phase to keep track of during the process is the minus sign arising from the qubit swap}. \\\\
In what follows, we deal with both protocols of Theorem $3$ in a subtle way. This will appear clear later in the discussion. Suppose the interferometric measurement outcome is $2$ instead of $0$ or $4$. Then, follow process (P) of \cite{CL2}, that is do the following braid on the last three anyons
\begin{center}
\epsfig{file=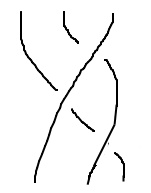, height=5cm}
\end{center}
We recall from \cite{BL} or \cite{CL} the matrix of this action on the qutrit $2222$. Up to overall phase, it is
$$\begin{array}{cc}&\begin{array}{ccc}|0>&|2>&|4>\end{array}\\&\\\begin{array}{l}|0>\\\\|2>\\\\|4>\end{array}&\begin{pmatrix} \frac{1}{2}&\frac{1}{\sqrt{2}}&\frac{1}{2}\\&&\\\frac{1}{\sqrt{2}}&0&-\frac{1}{\sqrt{2}}\\&&\\\frac{1}{2}&-\frac{1}{\sqrt{2}}&\frac{1}{2}\end{pmatrix}
\end{array}$$
Then measure the last two anyons.\\

If the outcome is $0$, after removing the pair of $2$'s to the right, we are back to the configuration we had before the interferometric measurement. Then, we bring another pair of $2$'s and run the interferometric measurement again. We are back to a previous time in history, namely to point $(t)$ like indicated on the two protocols. Up to point $(t)$ the two protocols are identical, hence we may still produce one gate or the other.\\

If the outcome is $4$, we fuse the last three anyons two by two, then fuse a pair of $4$'s in the last two anyons, using Lemma $1$ of Chapter $1$. Again, we are back to the configuration $(t)$ before the interferometric measurement and proceed from there. Again, we may still produce one gate or the other.\\

Finally, if the outcome is $2$, we have projected onto $|0>$ and $|4>$. In that case, we will produce the second gate of the Theorem. The protocol to finish is the following. Fuse the last three anyons two by two from the right hand side. We must keep track of a relative phase of $\pi$ between the $|0>$ and the $|4>$ as a result of the braids and the projection. Now do the qubit swap in order to switch the $|0>$ into a $|2>$ and switch the $|4>$ into a $|2>$.
The two minus signs, the one arising from the projection and the one arising from the qubit swap compensate beautifully.
We next finish like before by fusion followed by a $\sigma_1$-braid on the last two anyons. Again the relative phase of $-i$ introduced during the last braiding operation compensates exactly the relative phase of $i$ from the matrices (M). \\\\
\textbf{Thus, this time, the relative phases to keep track are:}\\\\
\indent\textbf{1) the minus sign when the right input bits are distinct}\\
\indent\textbf{2) the minus sign from the qubit swap, the one occurring right before point $(t)$}.\\


\newtheorem{Important remark}{Important remark}
\newtheorem{Very important remark}{Very important remark}



\noindent Work is not yet over as in both cases, we must still get back to a $2$-qubit shape now that the right qubit input has vanished. Below, we drew the configuration we are at.
\vspace{-0.2cm}
\begin{center}
\vspace{-0.4cm}
\epsfig{file=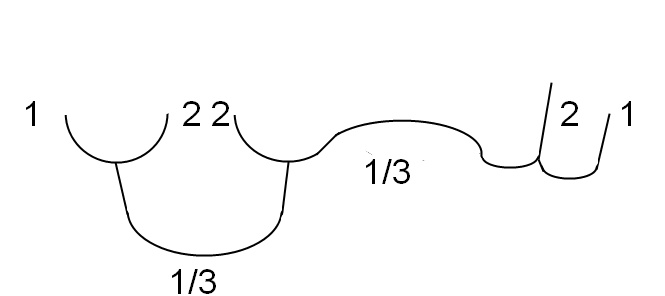, height=6cm}
\end{center}
The two edges carrying the $1/3$ superpositions will be the two edges carrying the $2$-qubit output. There will be two additional steps in order to reach the goal. First, we obviously don't have the right number of anyons. By unfusing the third anyon into two anyons of topological charge $2$, we increase this number and allow to get to the same topological shape as after the two inputs were initially fused. Except the central edge is a superposition of $1$ and $3$ from the F-move. \\\\\textbf{The F-symbols involved are opposite for distinct output bits. }
\hspace{-0.3cm}
\begin{center}
\epsfig{file=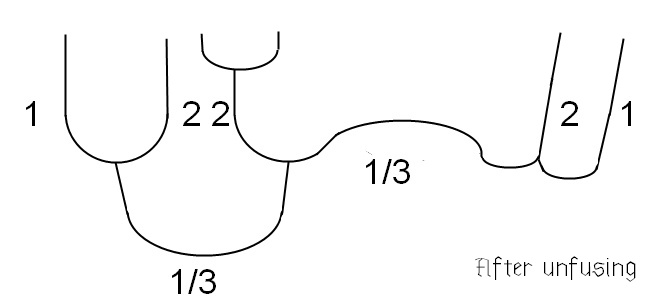, height=6cm}
\end{center}
\begin{center}
\epsfig{file=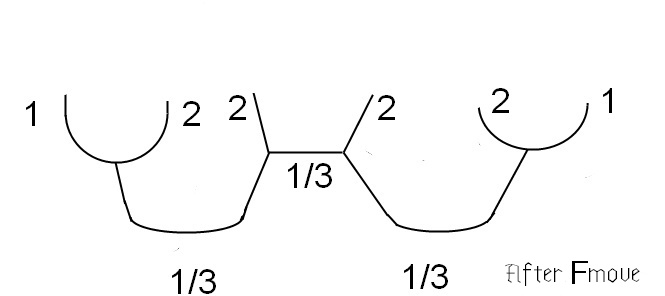
, height=6cm}
\end{center}

\noindent We force a label of either $1$ or $3$ of the central horizontal edge by running an interferometric measurement on the last three anyons.
If the resulting projection is onto $1$ (resp $3$), we next bring a pair of $1$'s (resp $3$'s) out of the vacuum and place it on top of the central horizontal edge. Then by running an interferometric measurement forced to $0$ of the last four anyons, we form a final "pair of $2$-qubit output". The goal is reached. If the outcome of the interferometric measurement is $2$ rather than $0$, we simply braid and remeasure, process we iterate until we have a successful outcome of $0$. Another way is to measure alternatively the last four anyons and the central two anyons until the separation occurs.\\

We are now able to conclude. In order to get the two entangling gates of the Theorem, we look back at the second entangling braid matrix and make the following changes. \\
\begin{itemize}
\item The last qubit digit of the rows which used to correspond to the right qubit input is no longer there. Therefore, we collapse the last four rows and the lower right quadrant of the matrix has to be moved up. \\
\item We take care of interchanging the two middle columns so that our input and output bases are ordered in the same and usual way. \\
\item Once this is done, the middle rows get a minus sign in order to take into account the third remark in bold. \\
\item Rows $2$ and $4$ get a minus sign because of the qubit swap occurring right before point $(t)$. \\
\item \textbf{For the second entangling gate only}, Columns $2$ and $4$ get a minus sign when the right qubit input is $3$ in order to take into account the second remark in bold.

    \end{itemize}
    $$\begin{array}{l}\end{array}$$
All together, we get the two entangling gates announced in Theorem $3$.

\begin{Important remark}
In the discussion above, we assumed that the outcome of the interferometric measurement of the three anyons at the top of the main figure of the Theorem is $1$ (just like on the figure of the Theorem). If the outcome is $3$ instead, the first and last rows of the gate get an overall minus sign (instead of the middle two). Anyhow, the two gates we get are the same since they simply differ by an overall minus sign. Also, it will suffice to fuse a pair of $4$'s into anyons $3$ and $4$ on one hand and $5$ and $6$ on the other hand in order to get back to a $2$-qubit $1221$. These two pair fusions operations get achieved at no cost.
\end{Important remark}

\begin{Very important remark}
The reader might wonder why we did the first full twist qubit swap. This is for good reasons which will be enlightened later. The way we operated all the different braids is the only way which makes the protocol of next section optimal in a sense we will define later.
\end{Very important remark}


\section{The CNOT. SWAP gate}

This section enlightens the fact that the $2$-qubit entangling gate which we made in \cite{CL2} is not only useful for universal quantum computation with the qubit $1221$ but is also a nice gate to have in its own. Together with our main protocol from previous section, we will use it to produce the product gate CNOT.SWAP; then we will deduce how to make the CNOT gate. In what follows, the auxiliary gate AUX is the one defined in $\S\,1$. The protocol producing it is the main protocol of \cite{CL2}. It is also the auxiliary gate used to make the circulant $2$-qubit entangling gate of \cite{CL2}.
\begin{Theorem}
By the main protocol of Theorem $3$, we are able to make one of the two gates $EG_1$ or $EG_2$. Then the product gate $CNOT.SWAP$ is given by
\epsfig{file=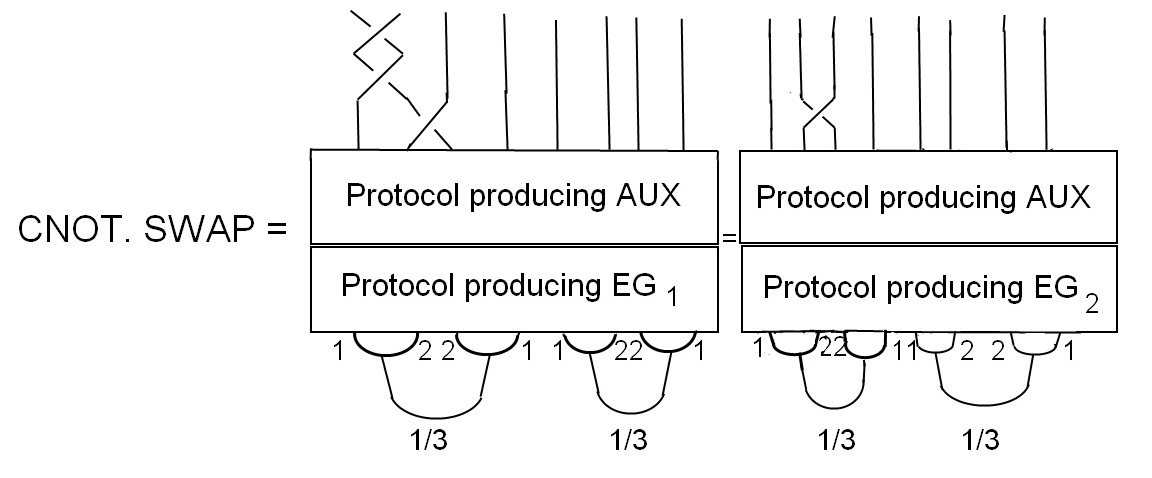, height=5cm}
\end{Theorem}
\textsc{Proof of Theorem $2$.} We have
$$AUX.EG_1=\begin{array}{cc}&\begin{array}{cccc}|11>\qquad&|13>\;&|31>\;&\;\;\;\;|33>\end{array}\\&\\\begin{array}{l}|11>\\\\|13>\\\\|31>\\\\|33>\end{array}&\begin{pmatrix}
-\frac{1}{2}\qquad&0\qquad&0\qquad&\frac{i\sqrt{3}}{2}\\&&&\\
0\qquad&\frac{i\sqrt{3}}{2}\qquad&-\frac{1}{2}\qquad&0\\&&&\\
-\frac{i\sqrt{3}}{2}\qquad&0\qquad&0\qquad&\frac{1}{2}\\&&&\\
0\qquad&\frac{1}{2}\qquad&-\frac{i\sqrt{3}}{2}\qquad&0
\end{pmatrix} \end{array}$$
and
$$AUX.EG_2=\begin{array}{cc}&\begin{array}{cccc}|11>\qquad&|13>\;&|31>\;&\;\;\;\;|33>\end{array}\\&\\\begin{array}{l}|11>\\\\|13>\\\\|31>\\\\|33>\end{array}&\begin{pmatrix}
-\frac{1}{2}\qquad&0\qquad&0\qquad&-\frac{i\sqrt{3}}{2}\\&&&\\
0\qquad&-\frac{i\sqrt{3}}{2}\qquad&-\frac{1}{2}\qquad&0\\&&&\\
-\frac{i\sqrt{3}}{2}\qquad&0\qquad&0\qquad&-\frac{1}{2}\\&&&\\
0\qquad&-\frac{1}{2}\qquad&-\frac{i\sqrt{3}}{2}\qquad&0
\end{pmatrix} \end{array}$$
Doing a $\sigma_2$-braid on the left input side 
yield the respective two matrices\\
$$\begin{array}{cc}&\begin{array}{cccc}|11>&|13>\;&\;|31>\;&\;|33>\end{array}\\&\\\begin{array}{l}|11>\\\\|13>\\\\|31>\\\\|33>\end{array}&\begin{pmatrix}
1\qquad&0\qquad&0\qquad&0\\&&&\\
0\qquad&0\qquad&1\qquad&0\\&&&\\
0\qquad& 0\qquad&0\qquad&-1\\&&&\\
0\qquad&-1\qquad&0\qquad&0
\end{pmatrix} \end{array}$$
$$\begin{array}{l}\end{array}$$
and
$$\begin{array}{cc}&\begin{array}{cccc}|11>&|13>\;&\;|31>\;&\;|33>\end{array}\\&\\\begin{array}{l}|11>\\\\|13>\\\\|31>\\\\|33>\end{array}&\begin{pmatrix}
1\qquad&0\qquad&0\qquad&0\\&&&\\
0\qquad&0\qquad&1\qquad&0\\&&&\\
0\qquad& 0\qquad&0\qquad&1\\&&&\\
0\qquad&1\qquad&0\qquad&0
\end{pmatrix} \end{array}$$
$$\begin{array}{l}\end{array}$$

\noindent It will now suffice to do a $\sigma_1$-full twist in order to swap the two $(-1)$'s of these matrices to $1$'s.
Indeed, the matrix for a $\sigma_1$-full twist is up to overall phase
$$\begin{array}{cc}&\begin{array}{cc}|1>&|3>\end{array}\\&\\\begin{array}{l}|1>\\\\|3>\end{array}&\begin{pmatrix}
1&&0\\&&\\0&&-1\end{pmatrix}\end{array}$$
After following the procedure of the figure, we get in each case
$$CNOT.SWAP=\begin{pmatrix}1&0&0&0\\0&0&1&0\\0&0&0&1\\0&1&0&0\end{pmatrix}$$

\begin{Very important remark}
To finish the section, we comment on Very important remark $1$. Let us introduce some notations. By operation $R_{ij}$ (resp $C_{ij}$), we mean that the rows (resp columns) $i$ and $j$ of a matrix get an overall minus sign. Once the second entangling braid matrix is put under the format of collapsing the rows to make the third bit vanish and exchanging the relative positions of the two middle columns, we note that the two operations we do are:\\

$R_{24}$ for the first entangling gate $EG_1$.\\
\indent $R_{24}$ and $C_{24}$ for the second entangling gate $EG_2$.\\

\noindent If we did other kinds of operations guided by different braiding schemes, we may need to act on an entangling gate twice by the auxiliary matrix instead of only once. This would be less efficient. For instance, if we did $C_{24}$ \textbf{only} (that would happen if we did not do the full twist prior to point $(t)$), that would be the case.
\end{Very important remark}

\section{All the $2$-qubit permutation gates}
In what follows, we will sometimes refer to "permutation gates" by the permutation that they encode. So, for instance the freshly made $CNOT.SWAP$ gate will be referred to as $(243)$. Because the SWAP gate is available to us by
\begin{center}
\epsfig{file=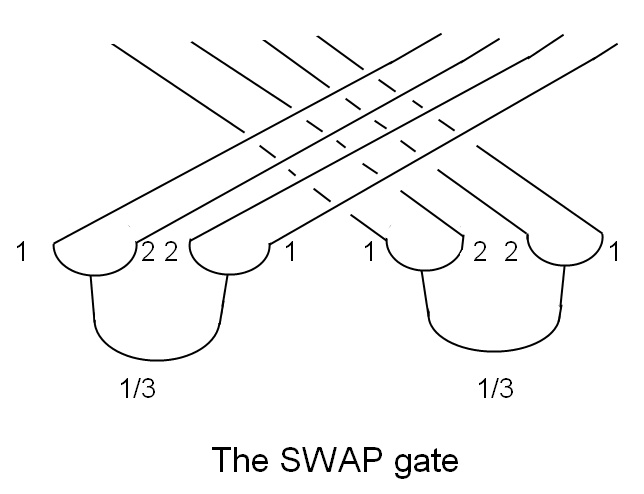, height=4cm}
\end{center}
we have the transposition $(23)$. And by Theorem $4$, we now have the controlled NOT gate which is the transposition $(34)$. Further, by simple pair fusion operations due to Mike Freedman \cite{CL}, thus called Freedman fusion operation (FFO), we also have all the products of two transpositions with disjoint supports, like follows.
\begin{center}
\epsfig{file=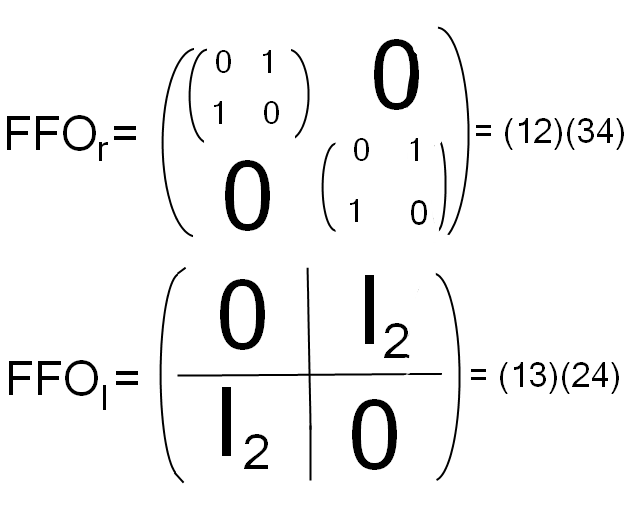, height=9cm}
\epsfig{file=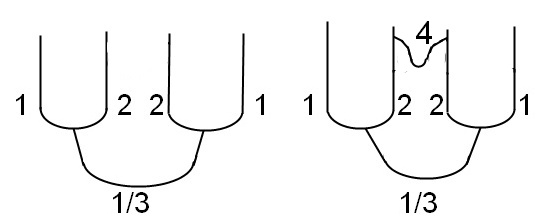, height=4cm}
\epsfig{file=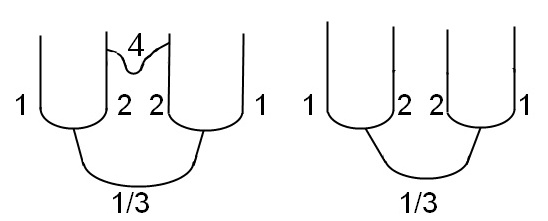, height=4cm}
\end{center}
The matrix product is the anti-identity matrix $(14)(23)$ and is obtained physically by realizing both fusions (on the right qubit and on the left qubit). \\
Multiplying a gate to the left by $FFO_r$ has the effect of swapping the first two rows and swapping the last two rows.\\
Multiplying a gate to the left by $FFO_l$ has the effect of swapping the quadrants up and the quadrants down. \\

Any $3$-cycle and product of two transpositions with disjoint supports suffice to generate $A_4$. Adding a transposition generator is enough to then generate the whole symmetric group $Sym(4)$. We thus obtain all the $2$-qubit permutation gates. \\

\noindent \textbf{Acknowledgements.} We thank Michael Freedman and Stephen Bigelow for helpful discussions.

\end{document}